\documentclass[article,preprint]{revtex4}
\usepackage{amsmath, amsthm, amssymb,graphicx,subfigure}

\begin{document}

\title{Interactions between Oxygen Interstitial and $\langle a \rangle$-Type Screw Dislocations in $\alpha$ Titanium}

\author{Liang Qi$^{1}$, Tomohito Tsuru$^2$,Mark Asta$^{3}$,D. C. Chrzan$^{3}$}
\email{dcchrzan@berkeley.edu}

\affiliation{$^1$Department of Materials Science and Engineering,
  University of Michigan, Ann Arbor, MI 48109, USA}

\affiliation{$^2$Nuclear Science and Engineering Center, Japan Atomic Energy Agency, 2-4 Shirakata-
Shirane, Tokai-mura, Ibaraki, Japan}

\affiliation{$^3$Department of Materials Science and Engineering,
  University of California, Berkeley, CA 94720, USA}

\date{\today}
\begin{abstract}
We employ density-functional-theory calculations to analyze the interactions between oxygen interstitial atoms and $\langle a \rangle$-type screw dislocations ($\langle a \rangle=\frac{a\langle11\bar{2}0\rangle}{3}$) in $\alpha$ titanium, based on investigations of generalized stacking fault (GSF) energies, dislocation core structures and the strain field of an oxygen interstitial. Compared with the substitutional atoms such as Al, there is a large repulsive interaction between oxygen interstitials at the octahedral site and an $\langle a \rangle$-type screw dislocation core, since the volume for the interstitial atom between Ti lattice sites is largely reduced in the core. Differential displacement maps and distributions of interstitial volume surrounding the dislocation core show that the interaction with an oxygen interstitial is very short ranged ($\sim$ 5 $\textup{\AA}$) and directional, mainly located on the prismatic plane where the majority of lattice displacements in the $\langle a \rangle$ screw dislocation core occurs. At long distances the interactions mediated by the strain from the interstitial oxygen and the dislocation vanish. We show that the large repulsive interaction at short range can push an oxygen atom from its original interstitial site to a new interstitial site on the basal plane in the dislocation core. The strong repulsive interaction and the mechanical shuffle of oxygen interstitials can be expected to have a large impact on dislocation dynamics, and can be used to explain the strong strengthening effects of oxygen impurities in  $\alpha$ titanium observed experimentally.
\end{abstract}
\maketitle

\section{Introduction}

Solid solution alloying and impurity atoms change the strength, plasticity and toughness of metallic alloys. Such effects are extremely strong for oxygen and nitrogen impurities in Ti and its alloys\cite{Conrad81}, which have many important structural applications, especially in aviation, aerospace and marine industries due to their high specific strength, excellent high-temperature mechanical performance and corrosion resistance\cite{Lutjering07}. For example, even an increment of 0.1 wt$\%$ oxygen impurities in commercial purity $\alpha$-Ti can significantly increase its yield strength dramatically, but with a concomitant reduction in toughness \cite{Lutjering07}. Recently in-situ transmission electron microscopy (TEM) compression tests were performed on Ti nanopillar samples with different oxygen concentrations\cite{Yu14}. The results show that only 0.3 wt$\%$ O can increase the yield strength to 2.5 GPa for $\alpha$-Ti . Further characterization and analyses indicate that such high strength mainly results from the interactions between oxygen interstitial atoms and screw dislocations with Burgers vector $\langle a \rangle=\frac{a\langle11\bar{2}0\rangle}{3}$ that glide on the prismatic planes $\lbrace 1\bar{1}00\rbrace$\cite{Yu14}. These recent findings have motivated the present computational work, aimed at further understanding the nature of the interactions between oxygen interstitial atoms and $\langle a \rangle$-type screw dislocation in $\alpha$-Ti.

Solid solution strengthening effects are generally regarded to originate from  two types of mechanisms, commonly referred to as elastic interactions and chemical interactions\cite{Conrad81,Yasi10}. The former is due to the long-ranged interactions between the strain fields of solute atoms and those due to dislocation lines\cite{Clouet08,Hanlumyuang10}. The latter can change the dislocation core structures and their mobility by varying the restoring force due to lattice slip; this type of interaction is short-ranged and can be related to the so-called generalized stacking fault (GSF) energy\cite{Vitek68,Joos97,Lu00,Yasi10}. The elastic interactions between solid solutions and the screw dislocation lines are relatively weak; according to isotropic linear elasticity theory, screw dislocations generate purely shear strain fields, which would lead to zero elastic interaction with solute atoms that generate only volumetric strains \cite{Hirth92}. Another type of elastic interactions is due to the change of elastic modulus because of solute atoms\cite{Conrad81}. However, such effect is not significant for 0.3 wt$\%$ O in $\alpha$-Ti lattice\cite{Kwasniak14}. Thus, the high strengthening effects for oxygen in $\alpha$-Ti more likely result from the chemical effects.

Using first-principles calculations, chemical effects underlying solid-solution strengthening can be investigated both indirectly and directly. Indirectly, first-principles calculations can be used to investigate the change of the GSF energy due to the presence of solute atoms on the slip plane, and the resulting effects on dislocation core structures can be estimated based on semidiscrete variational Peierls$-$Nabarro model\cite{Kwasniak13,Tsuru13}. Directly, the interaction between solute atoms and the core structure of a straight dislocation can be calculated either by the combination of lattice displacements distribution inside the core structure of pure metal and the change of GSF due to the solute atom\cite{Yasi10}, or by inserting the solute into the core structure in first-principles calculations of the core \cite{Trinkle05, Leyson10,Leyson12,Itakura13}. 

Short-ranged solute-dislocation chemical interactions can be classified into two categories depending on the lattice site for the solute atom. For substitutional solutes, replacing a matrix atom on a regular lattice site, the effects one the dislocation core may be relatively small as long as the solute has similar atomic radius compared with the matrix atom. Thus, it is accurate to estimate the chemical interaction based on lattice displacements inside the core structure of pure metals\cite{Yasi10,Leyson10,Leyson12}. On the other, the interaction between a dislocation core and an interstitial atom may be quite large especially when the latter have relatively large atomic radii; in such a case the interaction energy will depend on the interstitial volume, which could be quite different in the dislocation core than the bulk lattice, due to the large lattice displacements and distortions. For example, previously a model was proposed based on TEM characterization that $\langle a \rangle$ screw dislocation is slightly spread on the prismatic plane and would also dissociate into partial dislocations with $\langle c \rangle $-components on Burgers vector\cite{Sob75,Naka88}; these partial dislocations have to combine together for further slip on the prismatic plane (so-called \emph{sessile-glissile transformations}); the oxygen impurities could impede such re-combination processes to strengthen Ti\cite{Sob75,Naka88}. Recently several first-principles studies on the $\alpha$-Ti $\langle a \rangle$ screw dislocation core structure indeed show multiple complex core structures and there are still debates on the most stable configuration \cite{Tarrat09,Ghazisaeidi12,Tarrat14}.

To understand the origins of the significant strengthening effects of oxygen interstitials in $\alpha$-Ti, we employ in this work density functional theory (DFT) calculations of oxygen impurity effects on the GSF energy along the $\langle11\bar{2}0\rangle \lbrace 1\bar{1}00\rbrace $ slip system,  as well as on the $\langle a \rangle$ screw dislocation core structure in $\alpha$-Ti. We also calculate the lattice distortions due to oxygen interstitials in a bulk Ti lattice to estimate interactions with the strain field of screw dislocations, considering lattice anisotropy. Both GSF and dislocation core simulations show that, compared with the relatively weak strengthening effects of the substitutional solute atom Al, the extremely strong effect of oxygen and other large interstitial atoms result from the strong repulsion resulting from the large changes of interstitial volume inside the dislocation core. This strong interaction is found to be very directional (mainly on one prismatic plane) and short ranged ($<\sim$5 $\textup{\AA}$). In addition, the large repulsive interaction from the screw dislocation core can mechanically shift the oxygen from its original interstitial site to a new interstitial site on the basal plane in the dislocation core. We also discuss how the findings of the present atomistic calculations may affect the behavior of dislocation dynamics at larger scales. 

\section{Methods}
\subsection{Calculations of Generalized Stacking Fault Energy}
\label{sec_method_GSF}
DFT calculations were performed by using the Vienna ab initio simulation package (VASP)\cite{Kresse96a,Kresse96b}. All the calculations were performed non-spin-polarized, using the projector augmented wave (PAW) method \cite{Blochl94,Kresse99}, and the Perdew-Burke-Ernzerhof (PBE) exchange-correlation functional\cite{Perdew96}. There are 4 and 6 valence electrons for Ti (3$d^2$4$s^2$) and O (2$s^2$ 2$p^4$), respectively. Partial occupancies of eigenstates were determined by a first-order Methfessel-Paxton smearing scheme, with a broadening of $\sigma$ = 0.2 eV\cite{Methfessel89}. The cutoff energy for the plane wave basis was 400 eV.

\begin{figure}[th]
\subfigure[]{\includegraphics[width=0.35\textwidth]{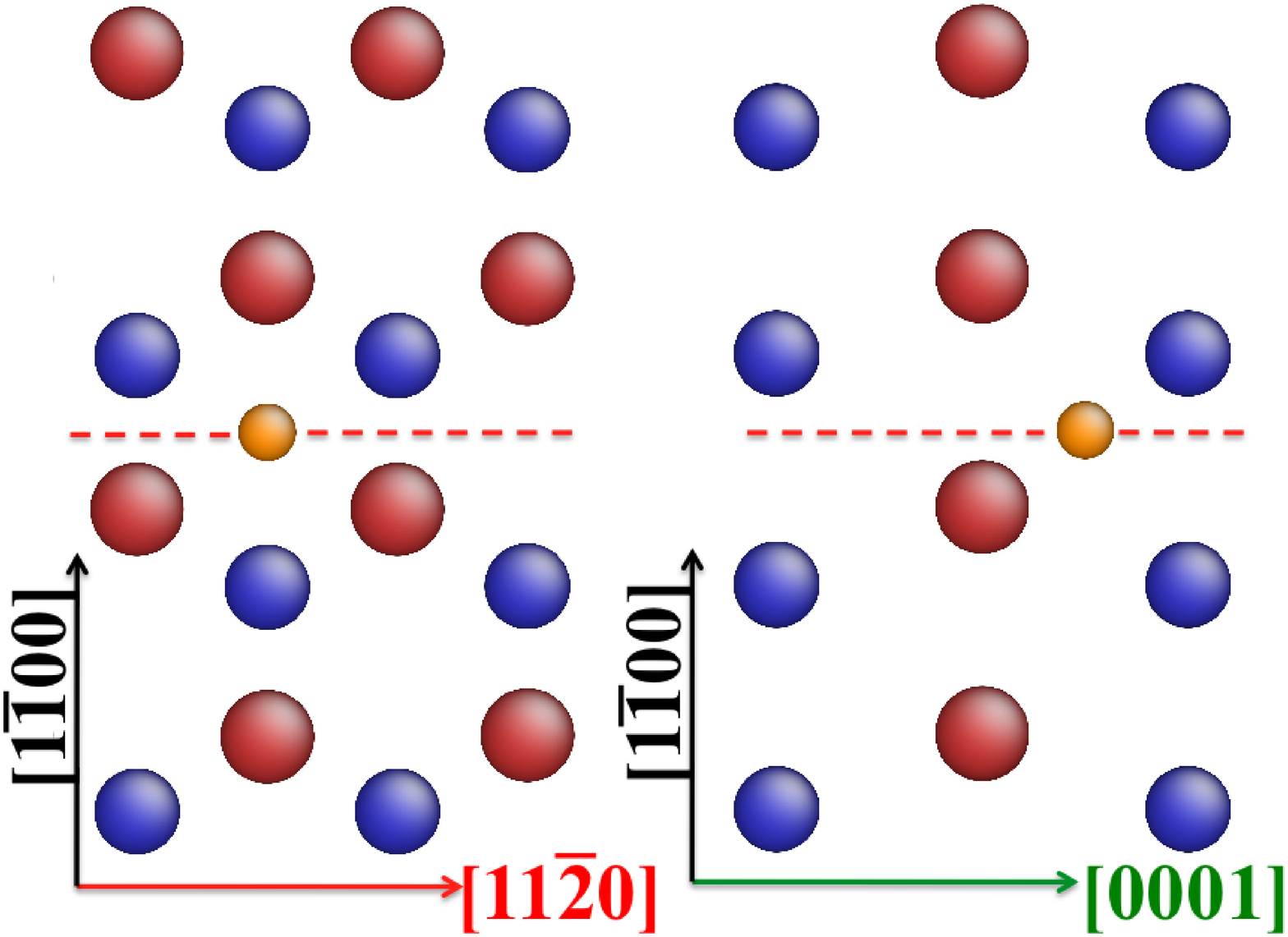}}
\subfigure[]{\includegraphics[width=0.55\textwidth]{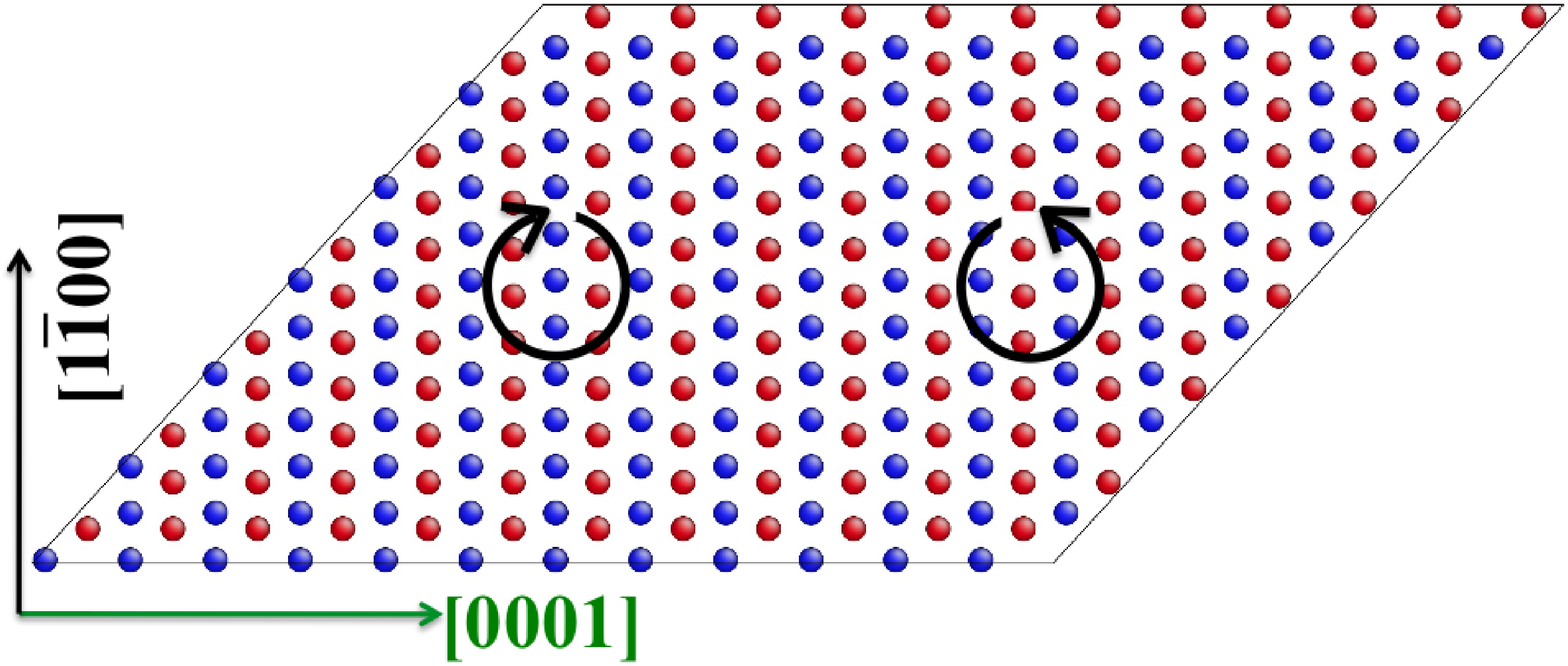}}
\caption{Supercells for (a) GSF and (b) dislocation core calculations. Ti atoms in type A/B basal planes are labeled as blue/red color. In (a) the dashed line is the location of slip interface, and oxygen atom is labeled as golden color on the interface. In (b) the clockwise and anti-clockwise circle stands for screw dislocation core with Burgers vector = $\pm\frac{[11\bar{2}0]}{3}$.}
 \label{fig_cells}
\end{figure}

As shown in Fig. \ref{fig_cells} (a), the supercells for the GSF energy calculations were constructed with a (2 $\langle a \rangle$ $\times$1 $\langle c \rangle$) periodicity within the slip plane, with the $x$-axis along the $\langle11\bar{2}0\rangle$ Burgers vector and the $y$-axis along $\langle0001\rangle$; along the perpendicular $z$-axis ($\langle1\bar{1}00\rangle$) there were 12 layers of prismatic planes plus 13 $\textup{\AA}$ of vacuum. For this supercell, (5$\times$7$\times$1) Monkhorst-Pack k-point grids for the Brillouin-zone integration were applied\cite{Monkhorst76}. An interstitial atom can be placed at the octahedral site on the interface, as indicated by the dashed line in Fig. \ref{fig_cells} (a), between two prismatic planes where the slip is set to occur. In the hcp structure there are two atoms in one primitive cell, one atom on type A and the other on the type B basal planes; these sites are indicated by blue and red colors, respectively, in what follows. There are two different geometrically-distinct types of prismatic interfaces. However, the one indicated by the dashed line in Fig. \ref{fig_cells} (a) is found to lead to much lower GSF energies, so that the other is not considered further in this work.

To understand solid-solution effects, a substitutional atom can replace a Ti atom on the prismatic plane nearest to the slip interface. Here substitutional atoms were used as a reference in order to compare the different strengthening effects from interstitial solutes. During GSF calculations, a fault vector $x$ was applied by shifting the top half of the supercell rigidly along the Burgers vector ($x$-axis) direction. Under these fault vectors, DFT calculations were performed by relaxing all metallic atoms only along the $z$-axis directionand the interstitial atom along all three axes. To check the effects of solid solution concentration, supercells with different sizes ((2 $\langle a \rangle$ $\times$2 $\langle c \rangle$) and (3 $\langle a \rangle$ $\times$2 $\langle c \rangle$)) were employed in calculations of the change of the GSF energy when one oxygen interstitial atom is placed on the slip plane (per area of the periodic supercell). The GSF energy results with oxygen interstitials depend on the initial position of the interstitial atom before DFT relaxations, so different initial oxygen positions were investigated.  Further details are given in Section \ref{sec_Results}.

\subsection{Calculations of Screw Dislocation Cores}
\label{sec_method_core}
We performed a series of DFT calculations of Ti screw dislocation cores without/with solute atoms. As shown in Fig. \ref{fig_cells} (b), the supercell has 12 $\langle c \rangle$ periodicity along the first vector [0001], 6 $ \langle a + c \rangle$ periodicity along the second [1$\bar{1}$01] vector, and 1 $\langle a \rangle$ periodicity along [11$\bar{2}$0] direction, so there are totally 288 Ti atoms in the supercell for pure $\alpha$-Ti. There are two straight screw dislocation lines along the [11$\bar{2}$0] direction with opposite Burgers vectors: $\frac{[11\bar{2}0]}{3}$ and $\frac{[\bar{1}\bar{1}20]}{3}$, respectively. Such a dislocation dipole configuration is called a quadrupolar stacking, where opposite-sign dislocations are stacked upon one another in a periodic dislocation dipole array to decrease the effects of long-range strain fields on dislocation core structures\cite{Cai03,Daw06}. The dislocation dipole configuration within the supercell also results in a distortion of the supercell vectors\cite{Lehto98}. Therefore, the lattice vector of the supercell along [1$\bar{1}$01] is tilted $\frac{\langle a \rangle}{2}$ along the Burgers vector direction to reflect a dipole composed of compact dislocations separated by one half the dimension of the supercell along the $\langle c \rangle$ direction. The initial displacement fields surrounding the dislocation core were solved algebraically using Daw's approach\cite{Daw06}, where the inputs are elastic constants of $\alpha$-Ti and the geometric centers of the dislocation cores. 

In setting up the initial displacement fields, Ti elastic constants were taken from experimental values, and we used two different geometric centers: one is at the octahedral site between two basal planes, the other is the center between two Ti atoms on the same basal planes. These initial structures result in slightly different dislocation core structures as observed in previous studies\cite{Tarrat09,Ghazisaeidi12,Tarrat14}, which will be discussed in the following sections. After relaxation to obtain the dislocation core structures in pure $\alpha$ Ti, Al as a substitutional solute and O as an interstitial solute were placed at various positions relative to the dislocation core centers. All atoms in the supercell were relaxed freely, but the supercell shape and volume were  constrained to be fixed. To save computational time, a ``soft" PAW potential for oxygen was used, so that the cutoff energy for the plane wave basis could be reduced to 282.8 eV in all dislocation core calculations.  A $1 \times 1 \times 11$ Monkhorst-Pack $k$-point grid was used for the Brillouin-zone integrations. The convergence criteria for the force was 0.01 eV/$\textup{\AA}$. All the other details associated with the DFT calculations were the same as described for the GSF calculations above. 

\subsection{Calculations of The Strain Field Induced by an Oxygen Interstitial}
\label{sec_method_strain}
The above calculations are used to calculate the interactions between solute atoms and screw dislocation cores at short distances ( $<$ 10 $\textup{\AA}$). At longer distances, the interactions between point defects and dislocations can be described by elastic theories based on the stress/strain fields of the dislocation and those of the individual point defect \cite{Eshelby56,Clouet08,Hanlumyuang10}. The strain field generated by $\langle a \rangle$ screw dislocation in $\alpha$-Ti can be obtained analytically by using the elastic constants of $\alpha$-Ti\cite{Chou62,Hirth92}. The strength of individual point defect can be described by the stress/strain of this point defect generated in a finite supercell from DFT calculations\cite{Hanlumyuang10}. Thus, we inserted one O interstitial atom at an octahedral site in a perfect Ti hcp structure, using an orthorhombic supercell (3 $\langle a \rangle$ along [11$\bar{2}$0], 2 $\langle c \rangle$ along [0001], 2 $\sqrt{3} a$ along [1$\bar{1}$01], respectively), and calculated the stress/strain due to this O impurity by using the same DFT methods for the GSF calculations. The O-induced stress was calculated by fixing the supercell and relaxing all the inner coordinates, and the stress-free strain was obtained by relaxing both the inner coordinates and the supercell shape/size. 

\section{Results and discussions}
\label{sec_Results}

\subsection{Effects of Solute Atoms on Generalized Stacking Fault Energies}
\label{sec_GSF}
\begin{figure}[th]
 \subfigure[]{\includegraphics[width=0.45\textwidth]{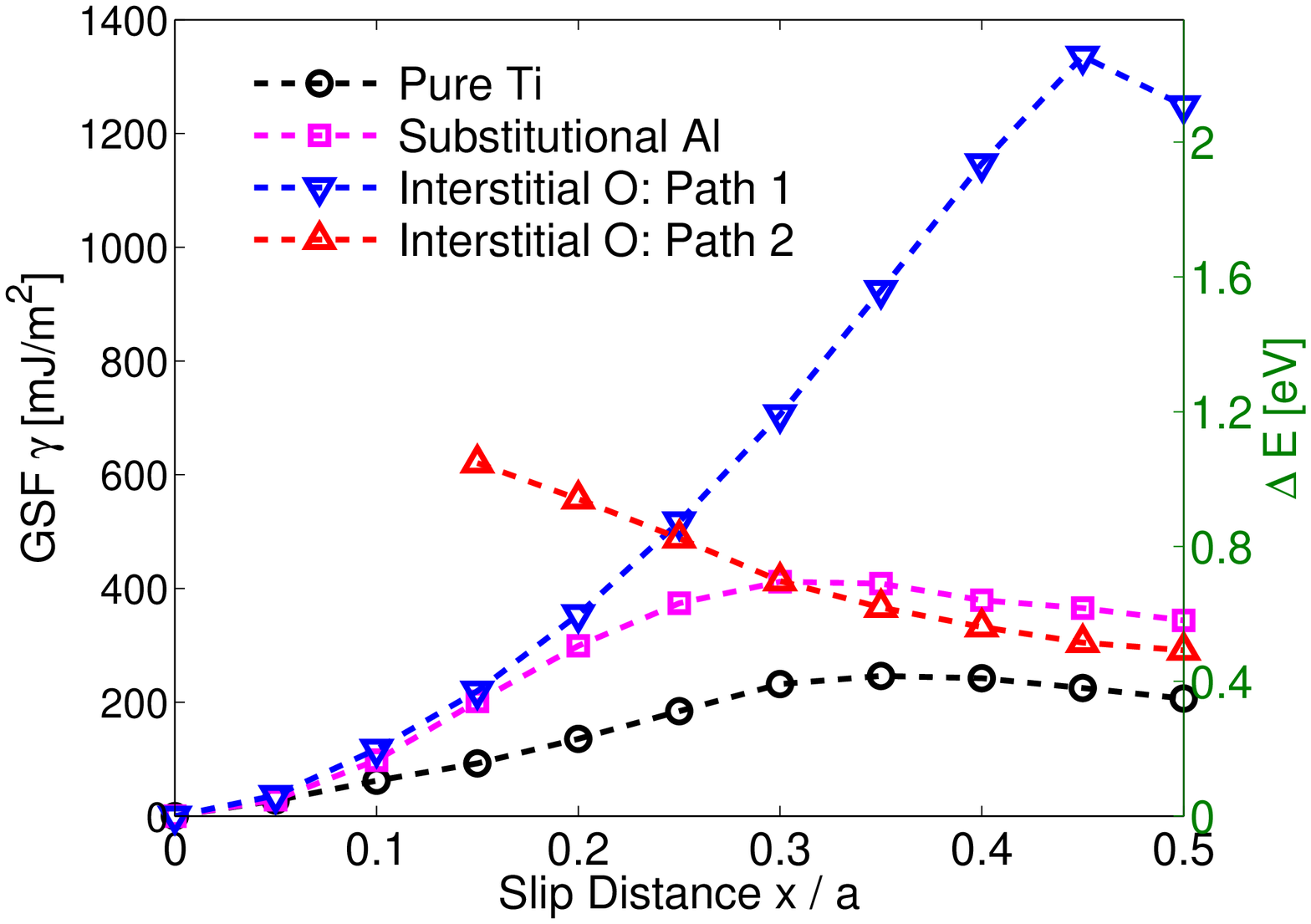}}
 \subfigure[]{\includegraphics[width=0.54\textwidth]{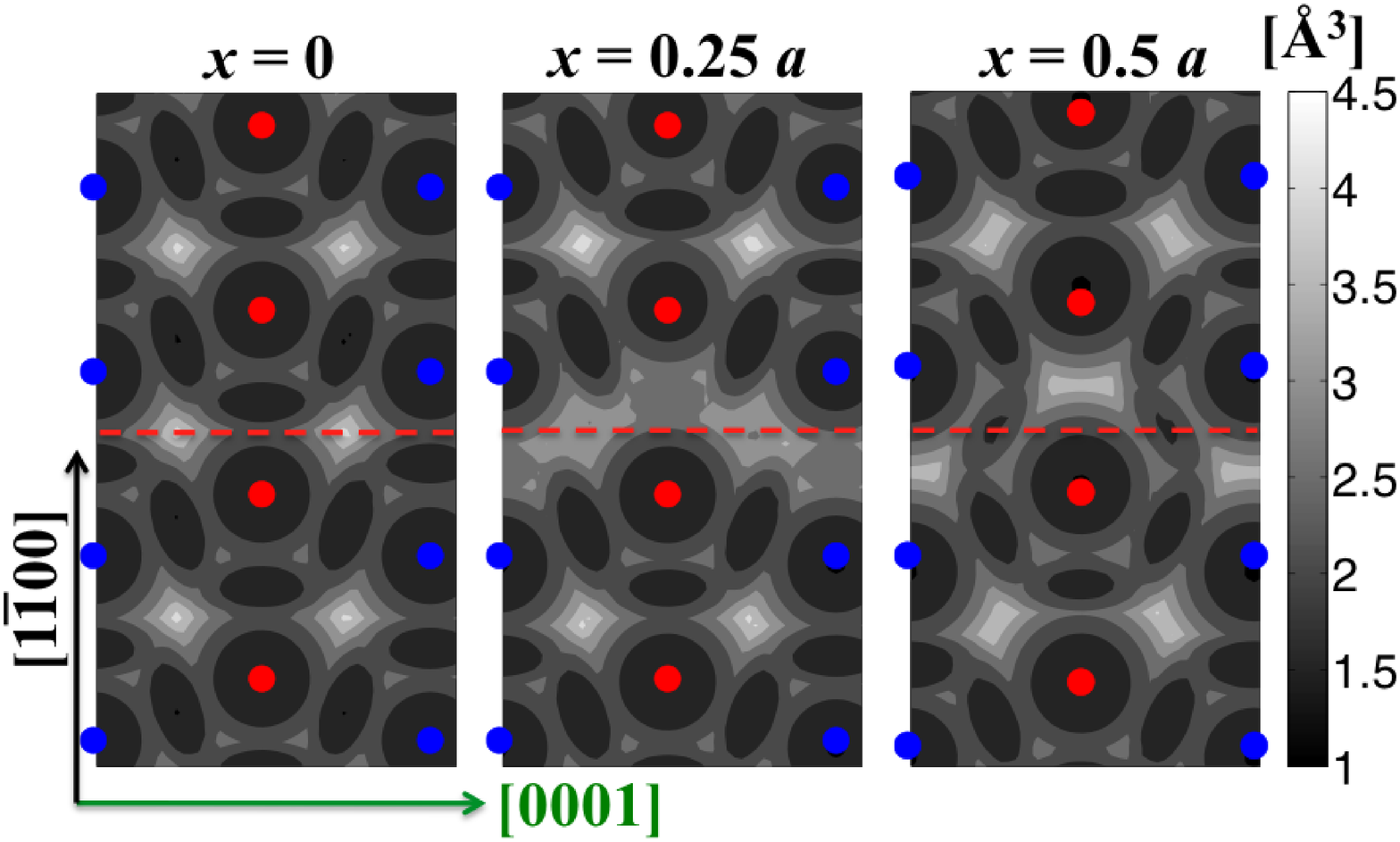}}
 \subfigure[]{\includegraphics[width=0.27\textwidth]{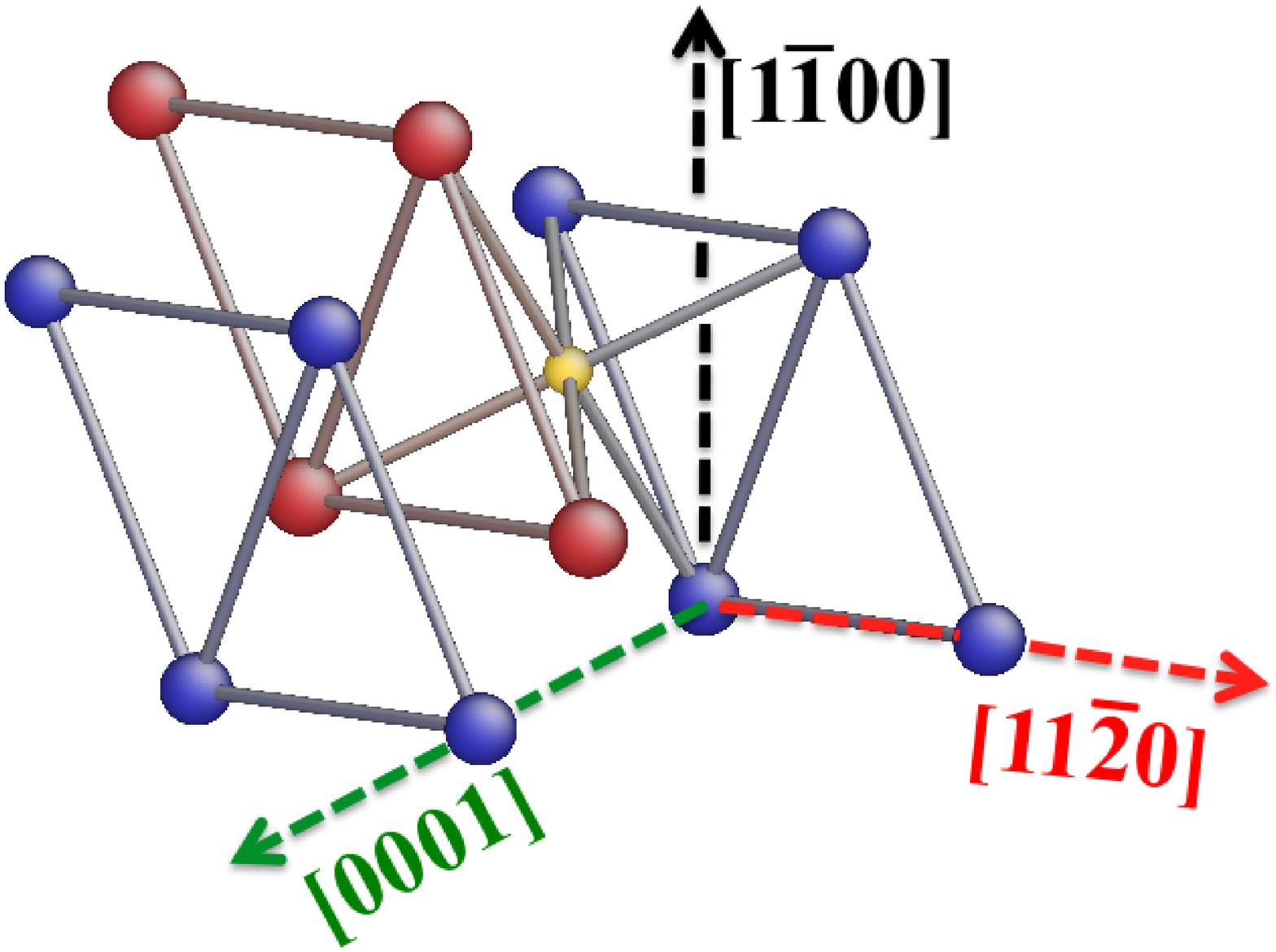}}
 \subfigure[]{\includegraphics[width=0.27\textwidth]{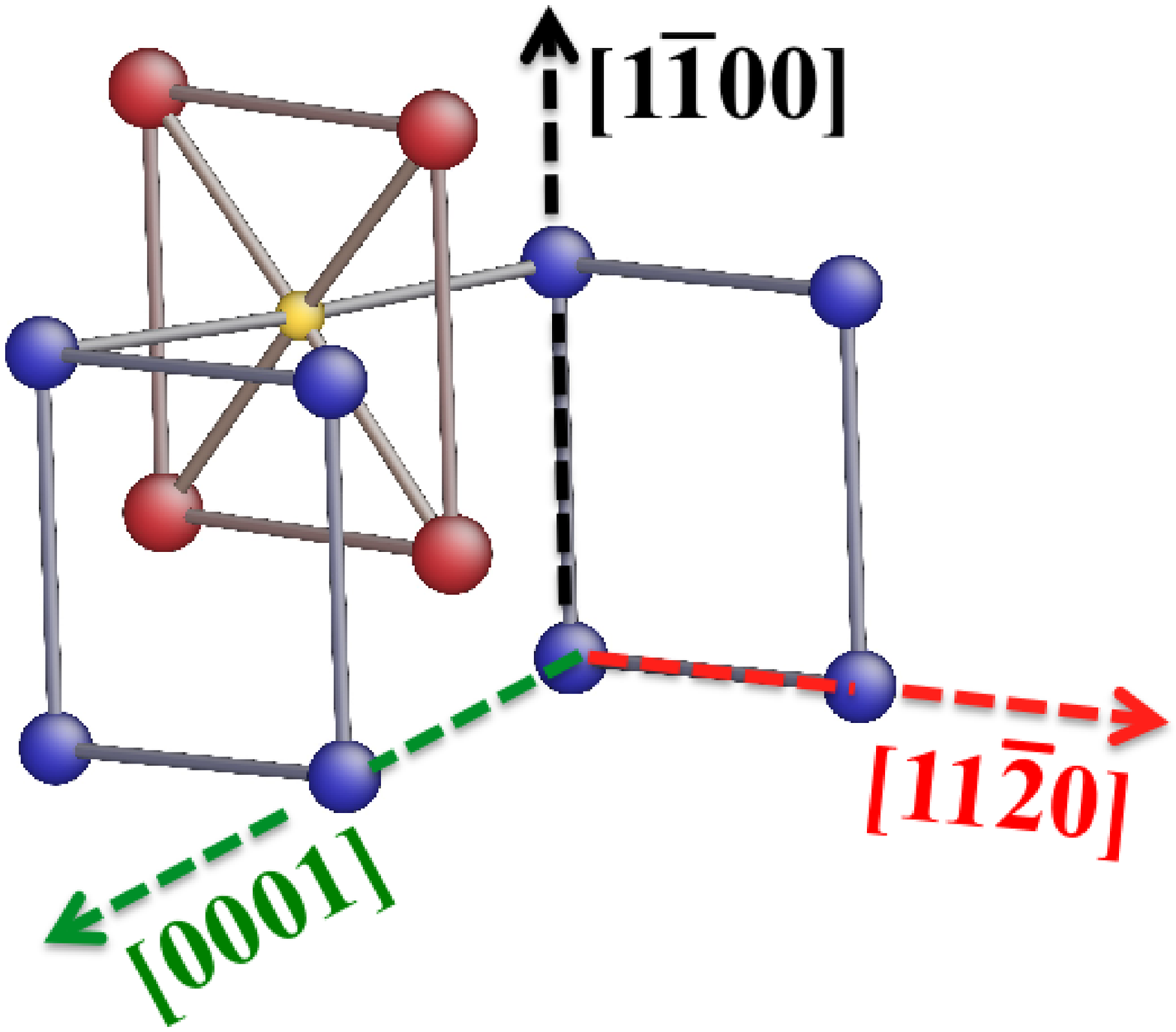}}
 \subfigure[]{\includegraphics[width=0.40\textwidth]{Fig2e}}
  \caption{(a) GSF for pure Ti and Ti with solid solution atoms on the interfaces. The left $y$-axis is labeled in unit of interfacial energy; the right $y$-axis is to show the corresponding energy difference in the (2 $\langle a \rangle$ $\times$1 $\langle c \rangle$) super cell. (b): Distribution of interstitial volume in $\alpha$-Ti lattice near the prismatic interface as the slip distance gradually increases. At any point, the maximum value of interstitial volume along $[11\bar{2}0]$ is plotted. (c): Lattice structures of oxygen (golden) at octahedral site between type A (blue) and type B (red) basal planes in perfect Ti lattice. (d): Lattice structures of oxygen at octahedral site on type B basal plane when slip = 0.5 $\langle a \rangle$. (e): The energy paths for oxygen shuffle from (c) (reaction coordinate = 0) to (d) (reaction coordinate = 1) at different fixed slip distance.}
 \label{fig_GSF}
\end{figure}

As determined by the lattice structure, the full GSF energy curve for $\alpha$-Ti on the prismatic plane for slip along $\langle a \rangle$ is symmetric with respect to the slip distance $x$ = 0.5 $a$, which corresponds to the maximum misfit across the prismatic interface. Thus we only show GSF results from $x$ = 0 to $x$ = 0.5 $a$ in Fig. \ref{fig_GSF} (a). For pure Ti, the GSF energy at $x$=0.5 $a$ takes the value $\gamma_{\textup{Ti}}(x=0.5a)$ = 206.7 mJ/m$^2$, and it corresponds to a shallow local minimum, consistent with previous studies\cite{Ghazisaeidi12}.

As mentioned in \ref{sec_method_GSF}, the GSF energy calculated with an interstitial atom on the slip plane was found to depend on the initial position of interstitial solute before relaxation. Since the stable site for the interstitial atom is determined by the available interstitial volume, we plot in Fig. \ref{fig_GSF} (b) the distribution of interstitial volume in the $\alpha$-Ti structure near the prismatic interface as the slip distance gradually increases. Here the interstitial volume of any point in the 3D lattice is defined as $\frac{4\pi}{3}d^3$, where $d$ is the distance between this point to its nearest Ti atom.  This definition is expected to provide a reasonable estimate of the space available to an interstitial atom, since the energy of the solute atom is expected to be strongly affected by its shortest Ti-interstitial bond length. We can see that in the perfect lattice ($x$ = 0) there are periodic regions with large interstitial volume located in the middle between a type-A and a type-B Ti atom on two nearby basal planes. Each of these regions corresponds to an octahedral interstitial site in Fig. \ref{fig_cells} (a).  The 3D atomic structure of oxygen at such an interstitial site is plotted in Fig. \ref{fig_GSF} (c).

As the slip distance increases, the interstitial volumes for the octahedral sites on the slip interface, labeled with a dashed line, gradually decrease. When the slip distance is equal to 0.5$a$, corresponding to the largest misfit across the slip interface, the interstitial volume near the original octahedral site on the slip interface is much less than half of the value in the perfect lattice. On the other hand, there are regions on the basal plane near the solid interface where the interstitial volume increases with increasing slip displacement, and they have almost the same interstitial volume as the original octahedral site in the perfect structure, when $x$=0.5$a$. As shown in Fig. \ref{fig_GSF} (d), this large interstitial volume corresponds to a new octahedral site on the basal plane when the slip distance close to 0.5$a$. The same conclusion regarding the transition of interstitial site was described recently in an independent study\cite{Ghazisaeidi14}.

Based on the above results, we set two paths for the GSF calculations when there is one oxygen interstitial atom on the prismatic interface. In Path I the oxygen is located at the original octahedral site between two basal planes before DFT relaxations, and oxygen in a nearby local minimum in the energy after relaxation. In Path II the oxygen interstitial is at the site that corresponds to the octahedral position on the basal plane for large slip distances, as illustrated in Fig. \ref{fig_GSF} (d). However, since this second site is not a local minimum in the energy for the undistorted Ti structure, the GSF energy for Path II was calculated starting from $x$= 0.5 $a$ with descending slip displacement.

Figure \ref{fig_GSF} (a) shows the calculated effect on the GSF energy due to solute atoms.  Results are plotted for O in both paths I and II described above, as well as for an Al substitutional atom on the prismatic interface. We first focus on the GSF energy with an interstitial O in Path I, which should be a natural choice for O starting from zero slip in the reference perfect lattice. Here both Al substitutional and O interstitial atoms increase the GSF energy along path I, as compared with results for pure Ti.  These results indicate that both solute species can strengthen Ti. In addition, the increase in GSF energy due to O along Path I are much higher than those from substitutional Al, for all ranges of slip distances. This suggests that the solid-solution strengthening effects of Al are weaker than for O. This strong effect of oxygen is easy to understand based on the large change of interstitial volume in Fig. \ref{fig_GSF} (b), which indicates that a very large strain is applied to the oxygen interstitial atom as the slip displacement increases. On the other hand, for the substitutional atom the change of atomic volume and the corresponding strain on the atoms are generally much smaller\cite{Yasi10,Leyson12}.

The calculated GSF energy values strongly depend on the solid-solution concentration on the interface, as determined by the supercell-size. Thus, it is necessary to obtain a quantitative value that is independent of the solid-solution concentration to describe the interaction between the dislocation core and an individual solute atom. We thus convert the value of the GSF energy, $\gamma$ per area, to a value per atom $\Delta E$ = $\gamma \times A_{\textup{s}}$, where $A_{\textup{s}}$ is the interfacial area in the supercell; this quantity is also labeled in the right $y$-axis in Fig. \ref{fig_GSF} (a). Further, we define the value $E^{\textup{int}}_{\textup{SA}}(x)$ as follows:
\begin{equation}
E^{\textup{int}}_{\textup{SA}}(x) = (\gamma_{\textup{SA}}(x)-\gamma_{\textup{Ti}}(x))\times A_{\textup{s}} = \Delta E_{\textup{SA}}(x) -\Delta E_{\textup{Ti}}(x)
\end{equation}
where $\gamma_{\textup{Ti}}(x)$ and $\gamma_{\textup{SA}}(x)$ is the GSF energy for pure Ti and Ti with a solute atom (SA) on the interface under the slip distance $x$, respectively. For supercells with large $A_{\textup{s}}$, $E^{\textup{int}}_{\textup{SA}}(x)$ is the energy difference to transfer a solid-solution atom from its equilibrium site in a bulk Ti structure to a site on the slip interface with slip distance $x$. Since in the center of a screw dislocation core there should be regions where $x$ = 0.5$a$, corresponds to the largest lattice misfit across the slip plane, $E^{\textup{int}}_{\textup{SA}}(x=0.5a)$ can be regarded as a reasonable estimate of the interaction energy between a single solute atom and the screw dislocation core.

The results in Fig. \ref{fig_GSF} (a) show that $E^{\textup{int}}_{\textup{O-Path I}}(x=0.5a)$ = 1.76 eV $\gg$ $E^{\textup{int}}_{\textup{Al}}(x=0.5a)$ = 0.23 eV, further confirming the strong repulsive interaction between the oxygen interstitial at the original octahedral site and the dislocation core, as compared with the smaller repulsive interaction for the substitutional atom. Increasing the supercell size for GSF calculations to (2 $\langle a \rangle$ $\times$2 $\langle c \rangle$) and (3 $\langle a \rangle$ $\times$2 $\langle c \rangle$),  $E^{\textup{int}}_{\textup{O-Path I}}(x=0.5a)$ increased to 1.94 eV, so this strong repulsive interaction is not a consequence of high solid solution concentration.  The magnitude of this repulsive interaction for interstitial oxygen is thus much stronger than the solid-solution effects associated with substitutional atoms, which are usually on the scale of 0.1 eV\cite{Yasi10,Leyson10,Leyson12}.

The GSF energy associated with an oxygen interstitial atom along Path II decreases as the slip increases. In the $2\langle a \rangle \times 1\langle c \rangle $ supercell, Path II becomes more favorable than Path I when $x$ $\geq$ 0.25$a$. This indicates that the energy can be reduced at high slip displacements by transferring oxygen from the original octahedral site, between two basal planes in the undistorted Ti structure, to the new octahedral site on the nearby basal plane. However, such an oxygen shuffle may require overcoming a high energy barrier, as oxygen diffusion in the bulk Ti structure is known to have barriers as high as 2 eV\cite{Wu10}. Thus, we estimate the barrier for the oxygen shuffle for two paths at fixed slip distance. Because the shuffle distance for oxygen is small and the shuffle path is simple, the barrier for oxygen shuffling at constant slip displacement can be estimated by linear interpolation of the atomic positions along the transition path, where the reaction coordination is the oxygen position along the $[0001]$ axis. In these energy-barrier calculations, for each intermediate step the oxygen coordination is fixed along the $[0001]$ direction but relaxed along the two other directions, and all metallic atoms are only relaxed along the z-axis to fix the slip distance

The results are shown in Fig. \ref{fig_GSF} (e). When the slip displacement is 0.50 $a$, oxygen at the octahedral site in Path II near the type B basal plane is 1.62 eV more stable than the site in Path I, and the energy barrier for shuffling between the two sites is 0.42 eV. When the slip displacement is 0.25 $a$, the stability of oxygen in the two paths is almost equal and the barrier for oxygen shuffling between the two sites is 0.95 eV. In addition, since there are two symmetry-equivalent directions for Path II, for oxygen near type A and type B basal planes, respectively, we investigate another oxygen shuffle path from Path I to Path II near a type A basal plane when the slip displacement is 0.30 $a$. The results show that oxygen at the octahedral site in Path II near the type A basal plane is 0.49 eV more stable than the site in Path I, and the energy barrier for shuffling between the two sites is 0.74 eV. So it can be estimated in the real dislocation core, the barrier for oxygen to shuffle between the sites in the two different GSF paths should be on the scale of 0.5$\sim$1.0 eV, which is comparable with the large oxygen diffusion barrier in the bulk Ti  structure\cite{Wu10}.  The large scale of the interaction energies and the energy barriers for oxygen shuffling (compared with $k_{\textup{B}}$ $\times$ 300 K = 0.0259 eV) suggest that oxygen (and most likely other similar large interstitials such as nitrogen) are strong dislocation pinning points. The large diffusion barriers also imply that the effects of dislocation slip and plastic deformation due to these pinning obstacles are likely to be sensitive to temperature and strain rate.

\subsection{Interactions between Solid Solution Atoms and Screw dislocations}
\label{sec_atom_core}

The above GSF calculations can be used to estimate indirectly the effects of solute atoms on dislocation core structures. However, in the real dislocation core, the slip distance and the corresponding lattice misfit is not a constant value but has an inhomogeneous distribution. In addition, there can be relative displacements between atoms other than those exactly located on the slip interfaces. To examine this possibility, we performed direct first-principles simulations of screw dislocation cores in $\alpha$-Ti, and study its interactions with solute atoms, as described in the next section.

\subsubsection{Dislocation cores in pure $\alpha$-Ti}
\label{sec_core_pure}

\begin{figure}[th]
 \subfigure[]{\includegraphics[width=0.45\textwidth]{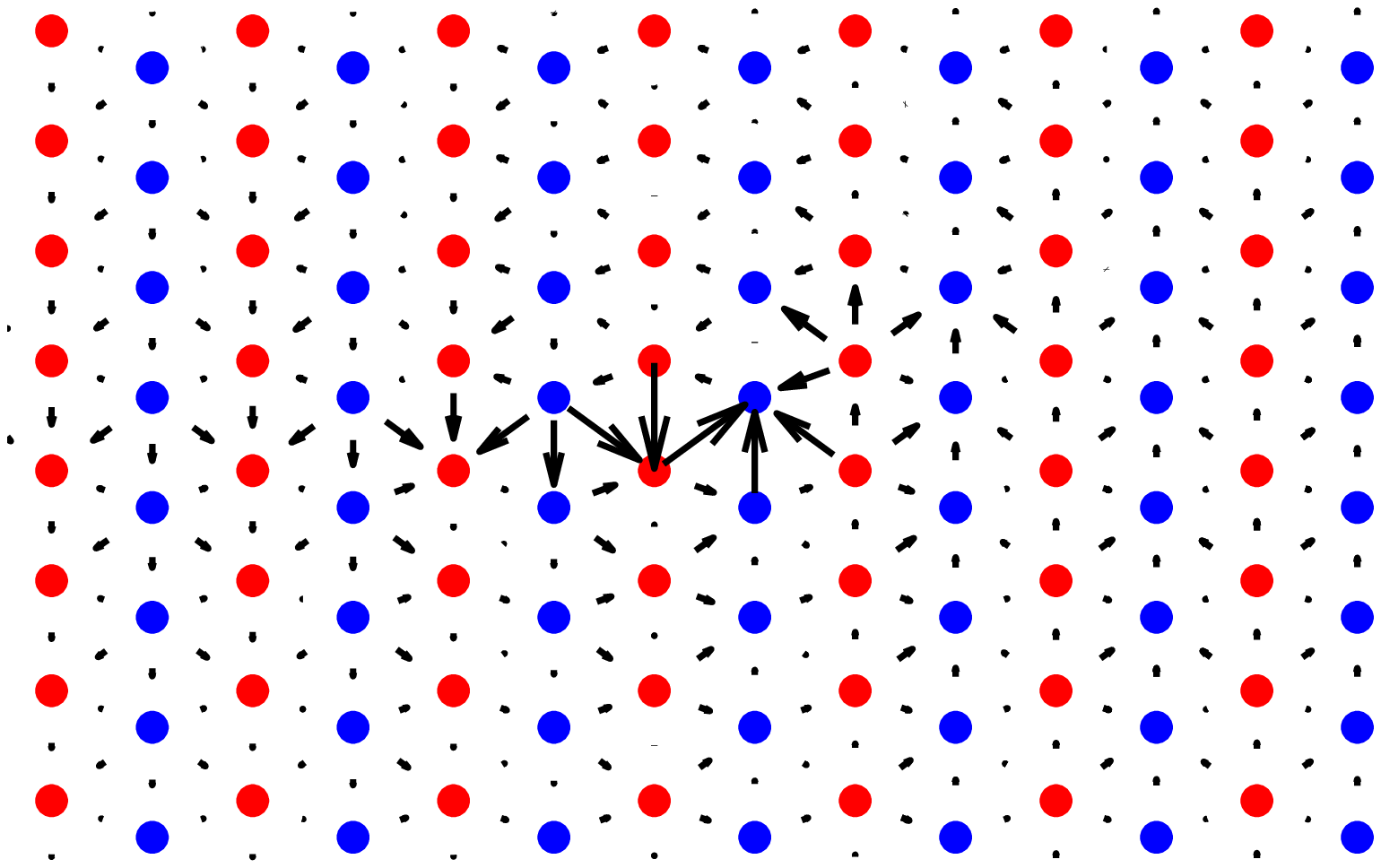}}
 \subfigure[]{\includegraphics[width=0.45\textwidth]{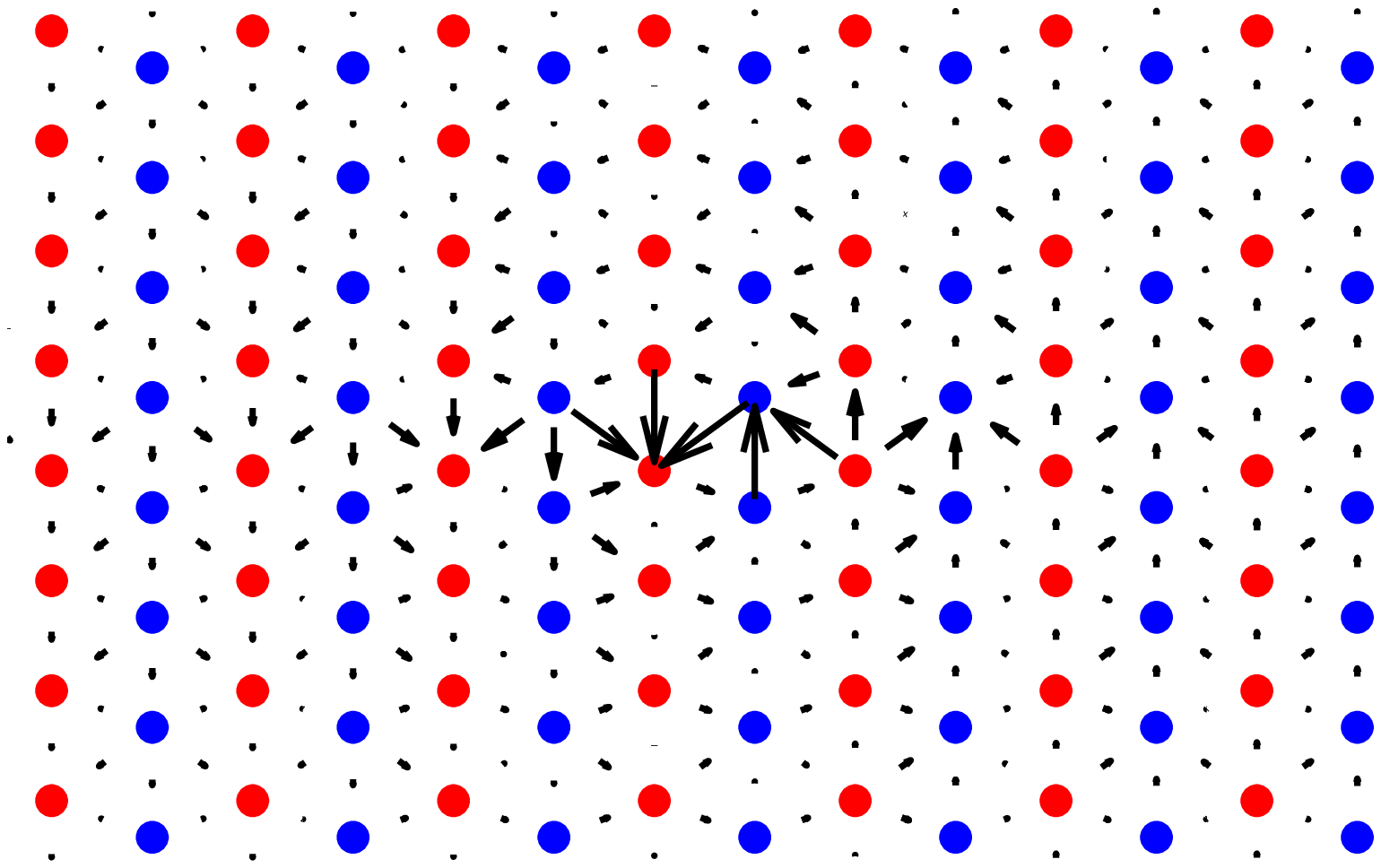}}
 \subfigure[]{\includegraphics[width=0.45\textwidth]{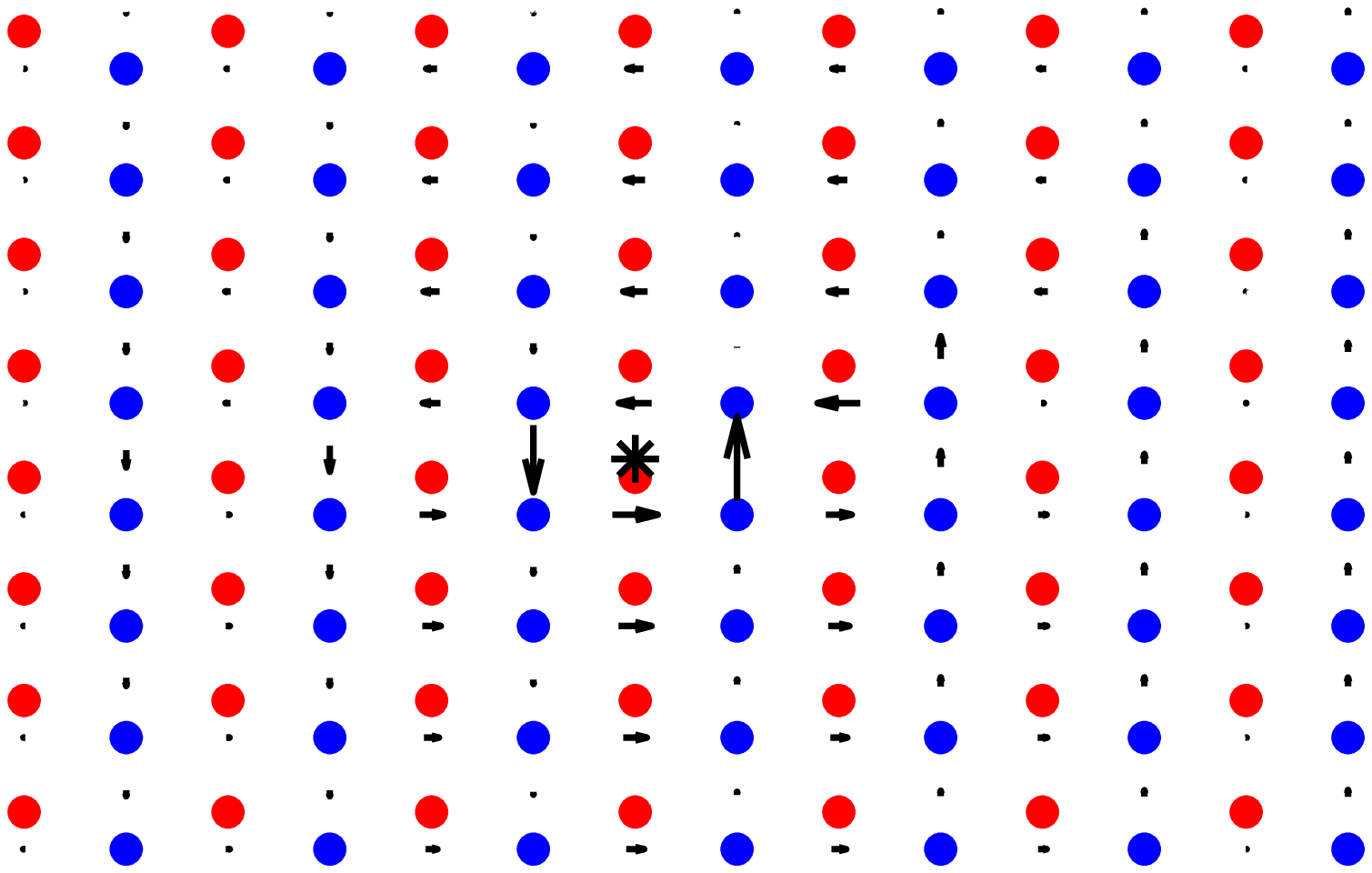}}
 \subfigure[]{\includegraphics[width=0.45\textwidth]{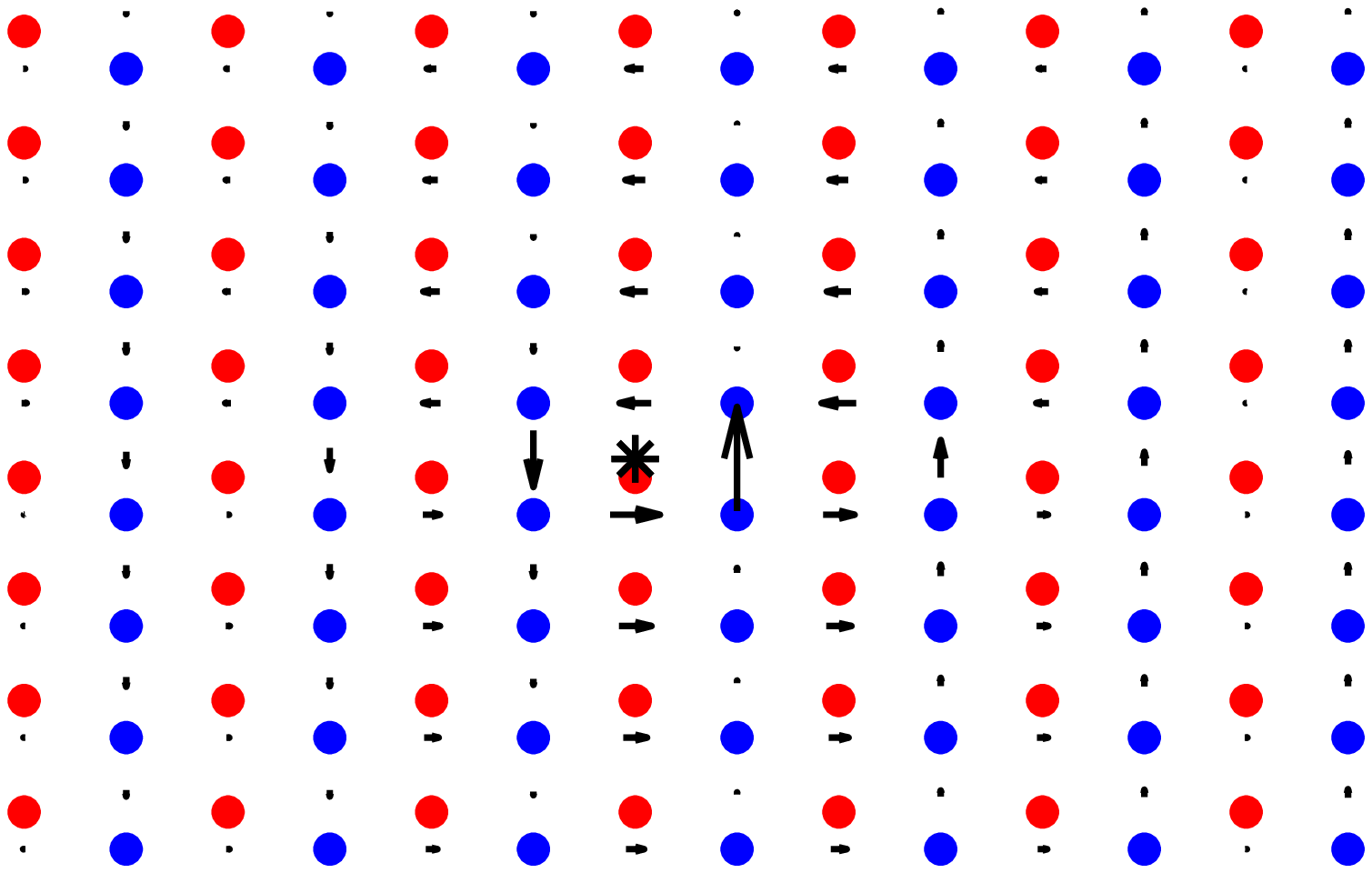}}
\subfigure[]{\includegraphics[width=0.45\textwidth]{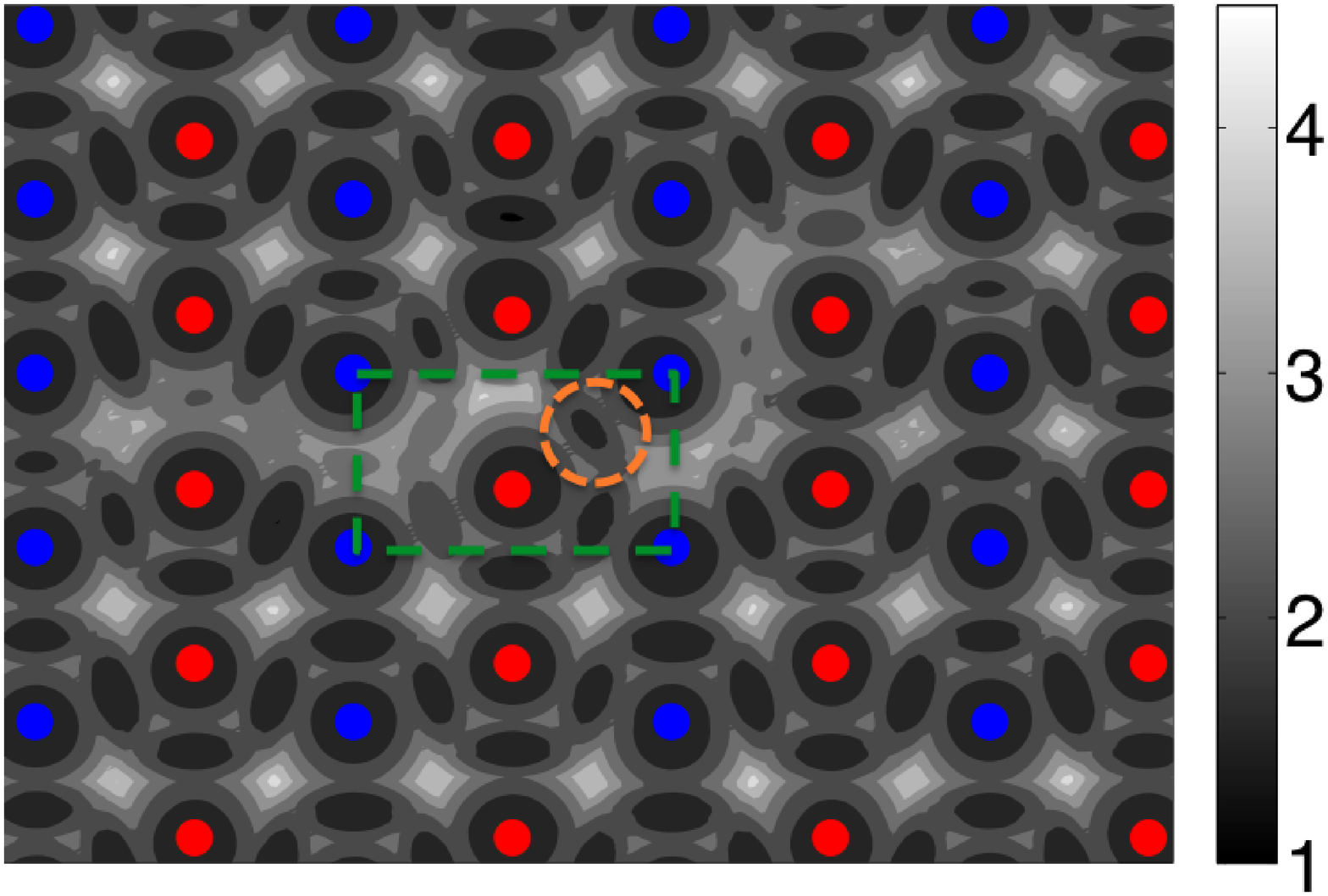}}
 \subfigure[]{\includegraphics[width=0.45\textwidth]{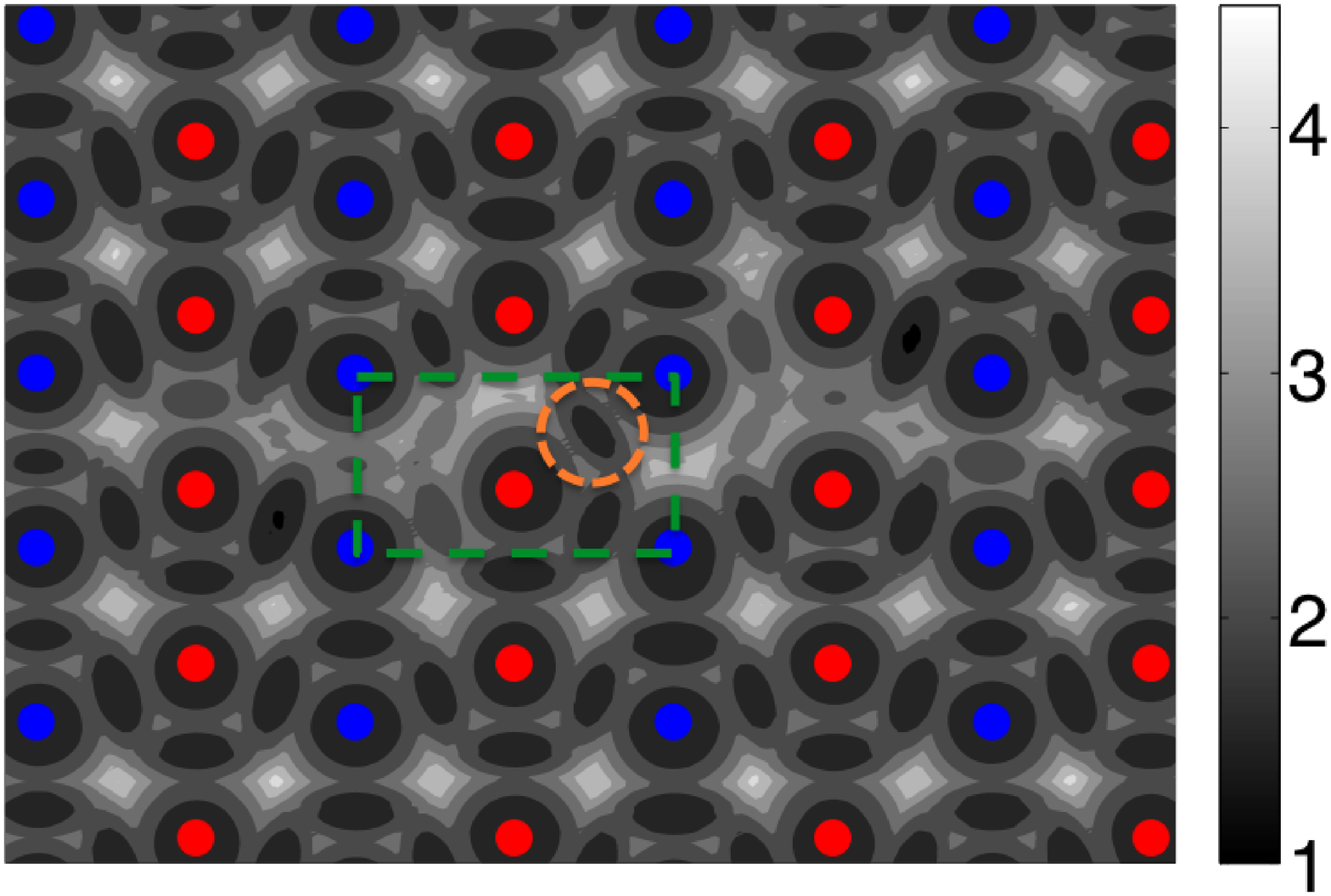}}
  \caption{(a) and (b): DD maps between each Ti atom for asymmetric core (a) and symmetric (b) core in pure Ti. The lattice orientation is the same as Fig. \ref{fig_cells} (b). (c) and (d): DD map between only type A (blue) Ti atoms for the same asymmetric core and symmetric core, respectively. Here * is located in the primitive cell with the integrated displacement equal to one Burgers vector. (e) and (f): Distribution of interstitial volume in the same asymmetric core and symmetric core, respectively. Dashed rectangles are located in the same primitive cell with * sign in (c) and (d). Dashed circles indicate the region that corresponds to an octahedral site in perfect lattice but with large reduction of interstitial volume in the dislocation cores.}
 \label{fig_core_pure}
\end{figure}

The relaxed configurations of the $\langle a \rangle$ screw dislocation cores in pure $\alpha$-Ti are plotted in Figure \ref{fig_core_pure} based on Vitek's differential displacement (DD) map\cite{Vitek70}. In these DD maps, the direction and magnitude of each vector between two nearest atoms stand for the relative displacement along the Burgers vector direction ($[11\bar{2}0]$, perpendicular to the plane of the paper) in the dislocation configuration compared with its counterpart in perfect lattice. In addition, the integration of the differential displacements surrounding a closed circuit on this map gives the total Burgers vector in the region surround this circuit. Here Fig. \ref{fig_core_pure} (a) is the DD map for the dislocation core configuration relaxed from the elastic solution where the geometric core center is located at the octahedral site between two basal planes; Fig. \ref{fig_core_pure} (b) is the map for the dislocation core relaxed from the elastic solution where the geometric core center is located on the basal plane. In both maps most of the displacement fields are spread on one prismatic plane along the [0001] direction, which is consistent with experimental reports that $\langle a \rangle$ screw dislocation mainly glides on a 2D prismatic plane\cite{Naka88, Yu14}. However, in Fig. \ref{fig_core_pure} (a) there are also relatively large displacements on another prismatic plane nearest to the glide plane. So Fig. \ref{fig_core_pure} (a) an (b) are commonly referred to as the asymmetric core and symmetric core, respectively\cite{Tarrat09,Ghazisaeidi12}. Thus, a slight difference of initial configuration determined by the geometric core center in the elastic solution results in different dislocation core structures. In our calculations under the given supercell structure, the asymmetric core is about 0.025 eV per Burgers vector more stable than the symmetric core, roughly the equivalent of room temperature. Although the energy differences may depend on the external strain field that varies with supercell structures, this small energy difference suggests that there are multiple possible metastable states of dislocation core configurations, and their energy differences are very small so that it is not difficult for the dislocation core structure to transform from one to the other\cite{Tarrat14}.

There are two atoms in each primitive cell in the hcp structure that can be labeled as atoms on type A and type B basal planes, denoted by blue and red colors, respectively. We plot another type of DD map to calculate the differential displacement vector only between the same types of atoms (either type A or type B), which can be used to describe the relative displacements between different primitive cells. Compared with DD maps between every atom, DD maps based on only type A/B atom can be easily used to identify the location of dislocation core by integration of displacement vector surrounding each primitive cell. For both the asymmetric core in Fig. \ref{fig_core_pure} (c) and symmetric core in Fig. \ref{fig_core_pure} (d), the integrated differential displacement surrounding a primitive cell formed by four nearest type-A basal atoms always show that the screw dislocation core can be located in just one primitive cell: the integrated displacement of the lattice rectangle surrounding the primitive cell is exactly one Burgers vector $\langle a \rangle$, but for all other primitive cells the integrated displacements are exactly zero. Same behaviors are also found for DD maps based on only type-B basal atoms for both cores. It means that even though the dislocation core structure is widely spread, it is still not dissociated into two partial dislocations; otherwise there should be at least two primitive cells that include finite Burgers vectors. We also increased the size of the supercell along the $\langle c \rangle$ direction by changing the periodicity of the supercell to 18 $\langle c \rangle$ $\times$ 4 $ \langle a + c \rangle$ and still did not observe dislocation dissociation.

Following the GSF calculations in Fig. \ref{fig_GSF} (b), we plot the distribution of interstitial volume surrounding the asymmetric and symmetric core in Fig. \ref{fig_core_pure} (e) and (f), respectively. In both figures, the green dashed rectangle is the primitive cell that contains integrated displacements equal to a full Burgers vector, as shown in Fig. \ref{fig_core_pure} (c) and (d). Consistent with the GSF results, in both dislocation core centers, there is more than a 50$\%$ reduction of the interstitial volume at the original octahedral site in the dislocation core, which is labeled by the dashed circle inside the primitive cell with full Burgers vector in Fig. \ref{fig_core_pure} (e) and (f); this is regarded as the dislocation core center in the pure Ti lattice in this paper. In addition, there are also large increases of interstitial volume on the two basal planes near this original octahedral site, where the new interstitial volumes on the basal planes are almost equal to those octahedral sites in the perfect lattice, consistent with the new interstitial site on the basal plane when slip = 0.5 $a$ in Fig. \ref{fig_GSF} (b) and (d). Thus, both the displacement maps and the changes of interstitial volumes confirm that there is a finite region in the center of the dislocation core with relative displacements very close to 0.5$a$. According to the GSF calculations, the repulsive energy between the dislocation core and an oxygen interstitial atom can be much larger than 1.0 eV, and such interaction can cause the oxygen to move to the new octahedral sites on the nearby basal planes.

\subsubsection{Solute atoms in the dislocation core}
\label{sec_atom_in_core}

\begin{figure}[th]
 \subfigure[]{\includegraphics[width=0.45\textwidth]{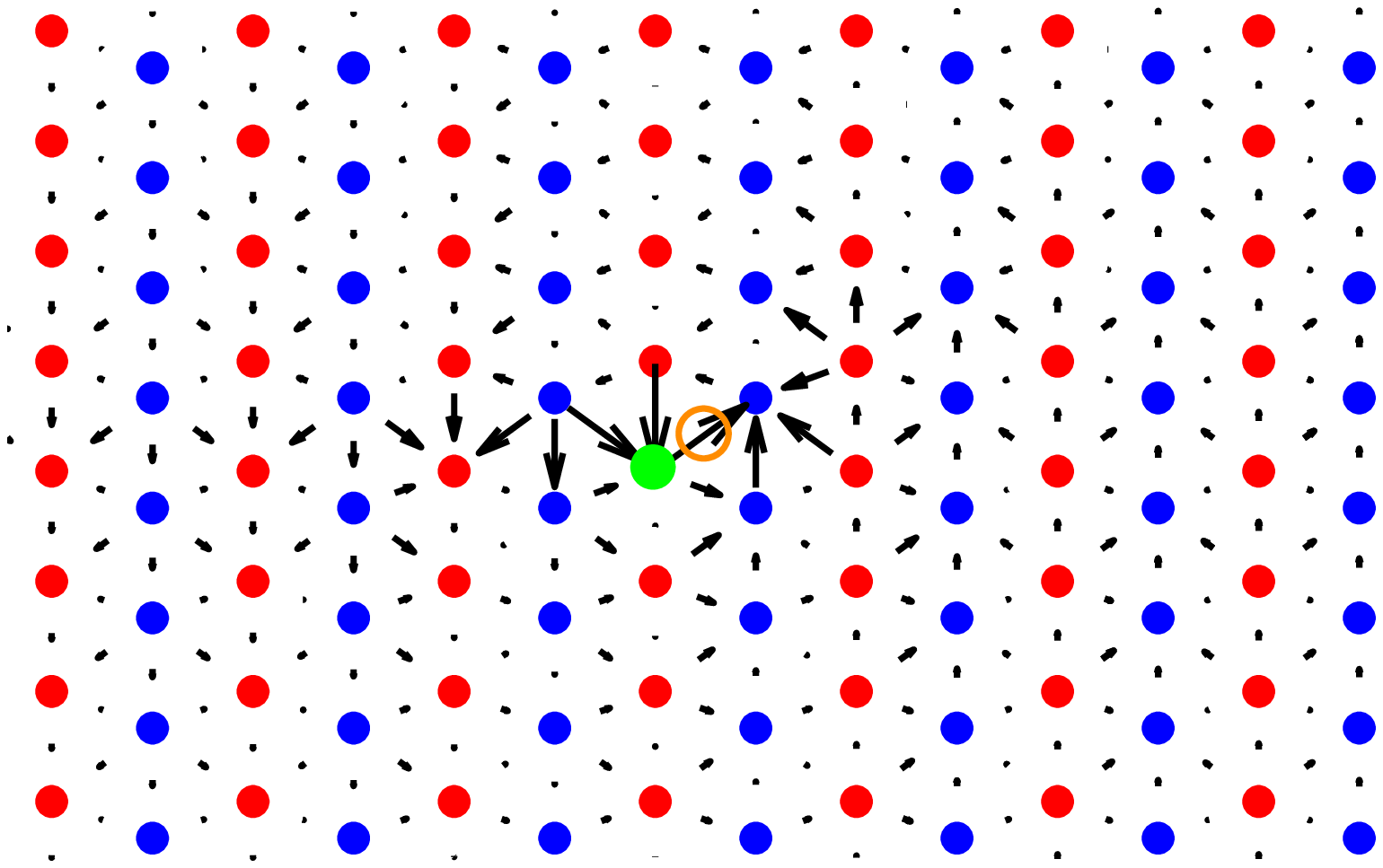}}
 \subfigure[]{\includegraphics[width=0.45\textwidth]{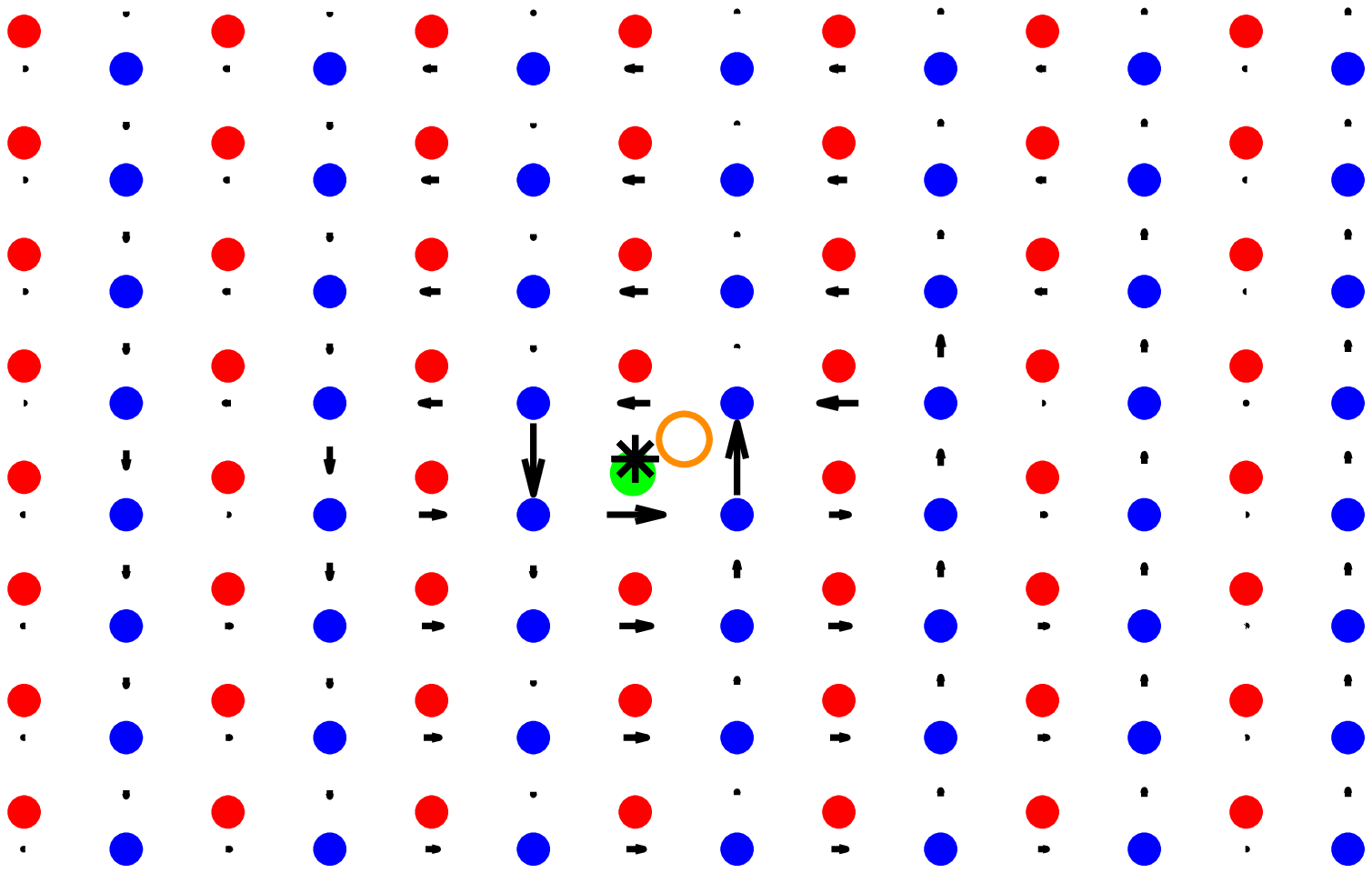}}
 \subfigure[]{\includegraphics[width=0.45\textwidth]{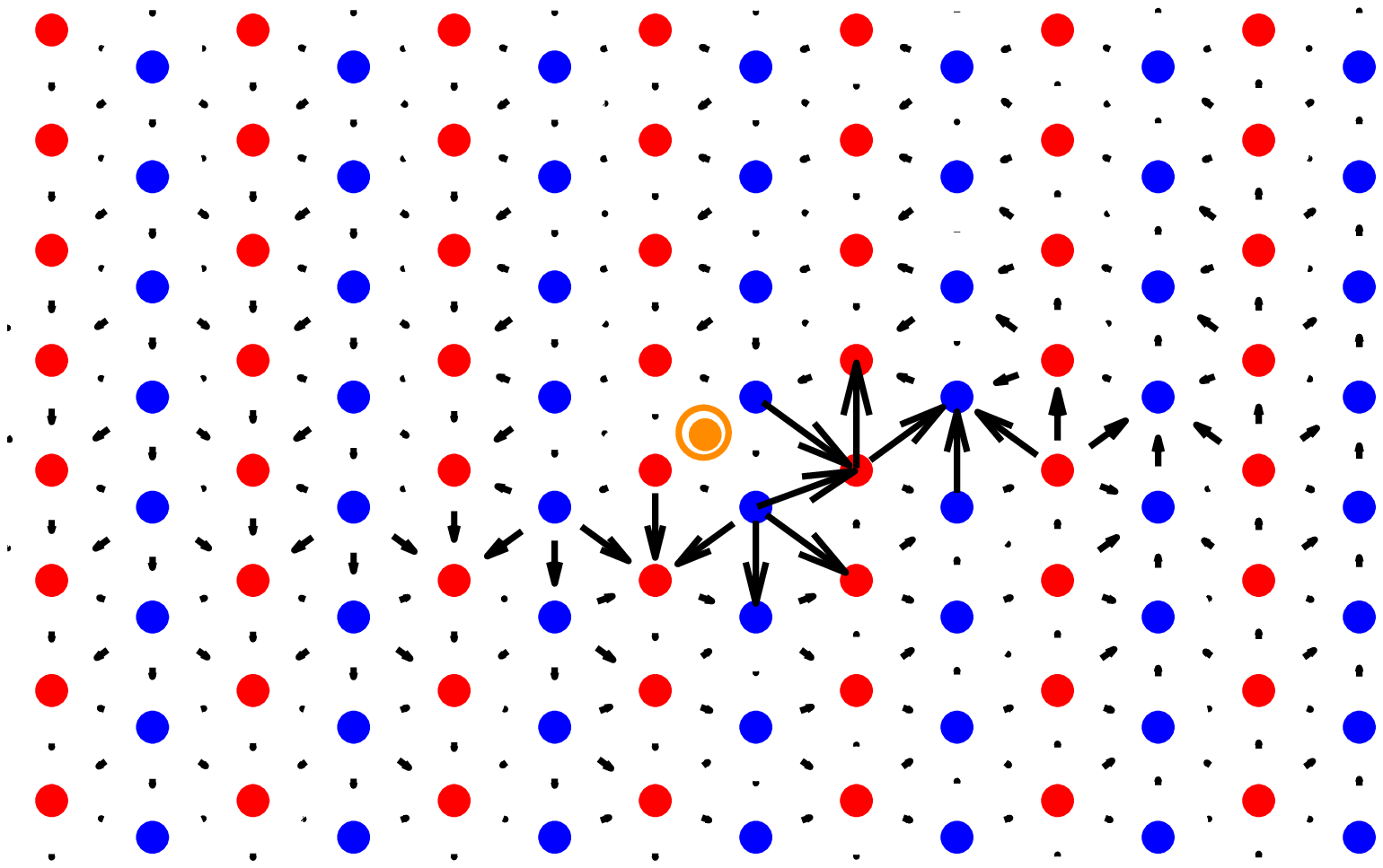}}
 \subfigure[]{\includegraphics[width=0.45\textwidth]{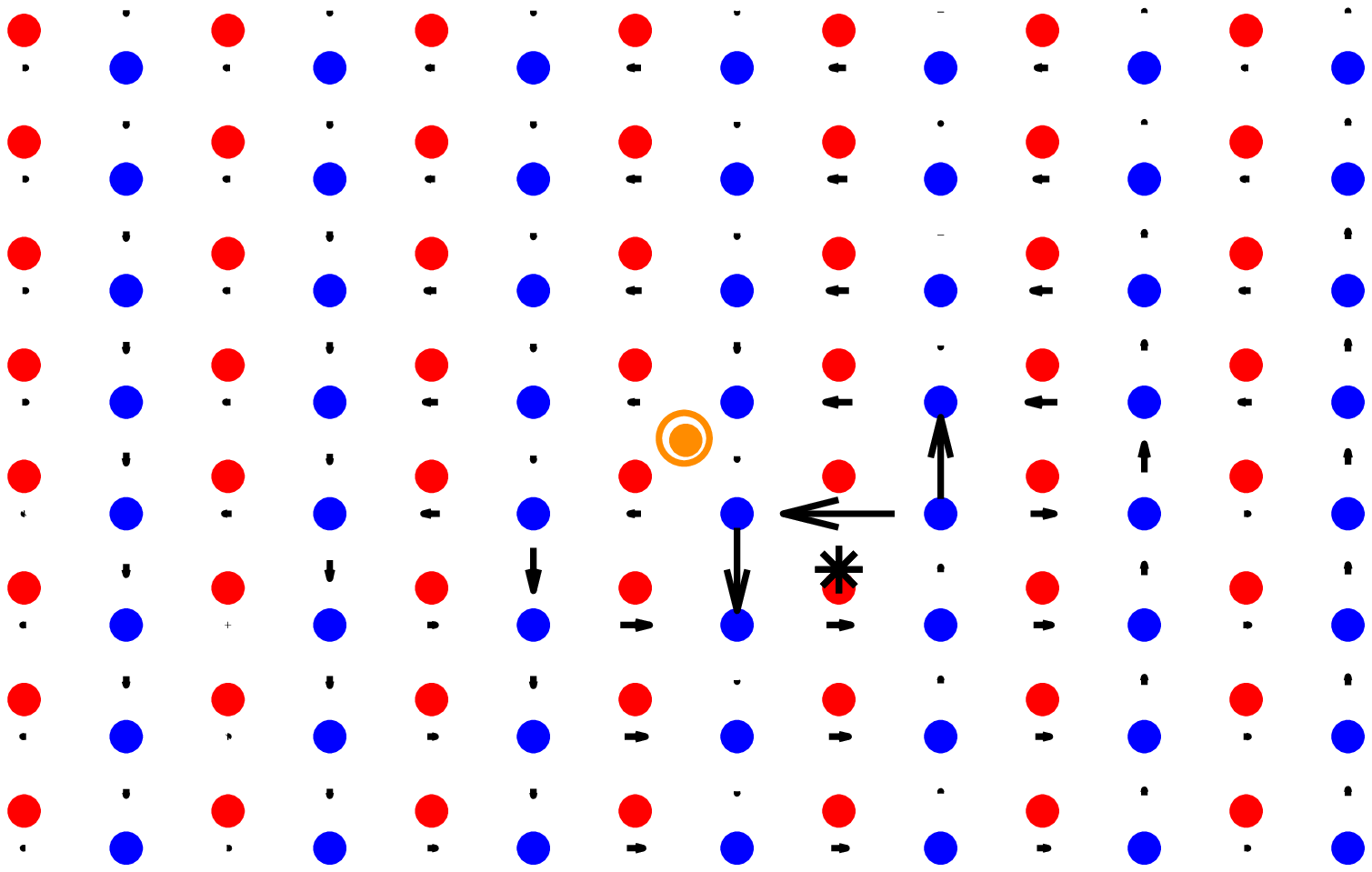}}
 \caption{(a) DD map between each Ti/Al atom when Al atom (green) is inserted in the asymmetric core. The golden circle is located in the same position as the dashed circle in Fig. \ref{fig_core_pure} (e). (b): DD map between only type A Ti atoms for (a). (c) DD map between each Ti atom when O atom (golden dot located in the golden circle) is inserted in the symmetric core. The circle is located in the same position as the dashed circle in Fig. \ref{fig_core_pure} (f). (d): DD map between only type A Ti atoms for (c). In both (b) and (d) * is located in the primitive cell with the integrated displacement equal to one Burgers vector.}
 \label{fig_core_ss}
\end{figure}

The next step is to study the interactions between solute atoms and $\langle a \rangle$ screw dislocation cores. First, for substitutional atoms, there are two Ti lattice sites on the two basal planes closest to the dislocation core center in pure Ti, indicated by dashed circle in Fig. \ref{fig_core_pure} (e) and (f). We replace a Ti atom in either of such two sites for both symmetric and asymmetric cores in pure Ti with Al atom and perform DFT relaxation calculations. Two types of DD maps with Al substitutional atom at such sites on type-B basal planes for the original asymmetric core are shown in Figure \ref{fig_core_ss} (a) and (b), where the filled green circle represents Al and the open golden circle is located at the same place of the dislocation core center in pure Ti indicated by dashed circle in Fig. \ref{fig_core_pure} (e). There are changes for each individual differential displacement vector by comparing Fig. \ref{fig_core_ss} (a) and Fig. \ref{fig_core_pure} (a),  The primitive cell which contains integrated displacements equal to a full Burgers vector is still located at the same position compared with the original DD map for pure Ti (Figure \ref{fig_core_ss} (b) vs. Fig. \ref{fig_core_pure} (c)), and there are also no other primitive cells with non-zero integrated displacements in Figure \ref{fig_core_ss} (b) as found also in pure Ti. This implies that the Al substitutional atom does not change the position of the dislocation core. We also obtain the same conclusion for Al at the other basal plane location (type-A basal) and in the symmetric core. The results are consistent with the finding that the repulsive interactions between substitutional solutes and the dislocation core structures are relatively small (on the scale of 0.1 eV for many alloying cases)\cite{Yasi10,Leyson10,Leyson12}.  It should be emphasized that the Al solute in these calculations has a high concentration because of small supercell dimension along $\langle a \rangle$; nevertheless, it is still found to be stable inside the dislocation core. This is consistent with the relatively small increase in the calculated GSF energies induced by Al substitutional atoms relative to pure Ti.

We also insert an oxygen interstitial into the dislocation core, which is located exactly in the position of dashed circle in Fig. \ref{fig_core_pure} (e) and (f). Two types of DD maps with O interstitials at such sites in the symmetric core are shown in Figure \ref{fig_core_ss} (c) and (d), where the golden filled circle represents the position of the O atom, inside the golden empty circle located at the same place of dashed circle in Fig. \ref{fig_core_pure} (f). As shown in Fig.\ref{fig_core_ss} (c), the DD map between each atom changes significantly from its original planar shape to a distribution on two nearby prismatic planes. Correspondingly, in Fig.\ref{fig_core_ss} (d) we can see that the primitive cell with one integrated Burgers vector is pushed away to anther prismatic plane. We find similar results for the case of the asymmetric core. These results imply that the repulsive energy between oxygen atom and the dislocation core is so large that the dislocation core is effectively pushed away by the interstitial solute. This is consistent with the large value of $E^{\textup{int}}_{\textup{O-Path I}}(x=0.5a)$, and the significant reduction of interstitial volume as slip increases in the GSF calculations.  On the other hand, the DFT relaxation does not induce the oxygen solute to move into adjacent sites, such as the new octahedral site on the basal plane in the dislocation core; this result is again consistent with the large activation barrier for an oxygen to shuffle to nearby site, as discussed in the previous section. 

Previous DFT simulations with different boundary conditions suggest that Ti [11$\bar{2}$0] screw dislocation core may dissociate into two partial dislocations because of the local minimum in GSF located at slip = 0.5 $a$ in Fig. \ref{fig_GSF} (a)\cite{Ghazisaeidi12}. However, if the dissociation exists, the region between two dissociated dislocation cores should be a stacking fault layer with relative slip = 0.5 $a$. According to our GSF calculations, this means that the region with small interstitial volume would further grow, and would increase the repulsive interaction between oxygen interstitial atoms and the dislocation core further.

\subsubsection{Solute atoms near dislocation cores}
\label{sec_atom_near_core}

To further clarify the magnitude and character of the strong repulsive interactions between an oxygen interstitial and a dislocation core, we also insert oxygen solutes at different interstitial sites. Two types of interstitial positions were investigated. One type is the octahedral site on the same glide prismatic plane of the dislocation core with distance $\pm \frac{\langle c \rangle}{2}$, $\pm\langle c \rangle$ and $\pm \frac{3\langle c \rangle}{2}$, respectively. The other type is on one of the two nearest prismatic planes closest to the glide prismatic plane of the dislocation core, and the distance between the core octahedral site to the investigated site along $[0001]$ is 0, $\pm \frac{\langle c \rangle}{2}$, $\pm\langle c \rangle$ and $\pm \frac{3\langle c \rangle}{2}$, respectively. In total, there are 20 octahedral sites investigated for asymmetric and symmetric cores. To calculate the interaction energies between oxygen interstitials at those sites and in the dislocation core, a reference state is also calculated by inserting one oxygen in the supercell at a distance $>$16 $\textup{\AA}$ away from the dislocation core, for both asymmetric and symmetric configurations. Based on this reference, the interaction energies between the oxygen interstitial and $\langle a \rangle$ screw dislocation core are calculated.

\begin{figure}[th]
\subfigure[]{\includegraphics[width=0.45\textwidth]{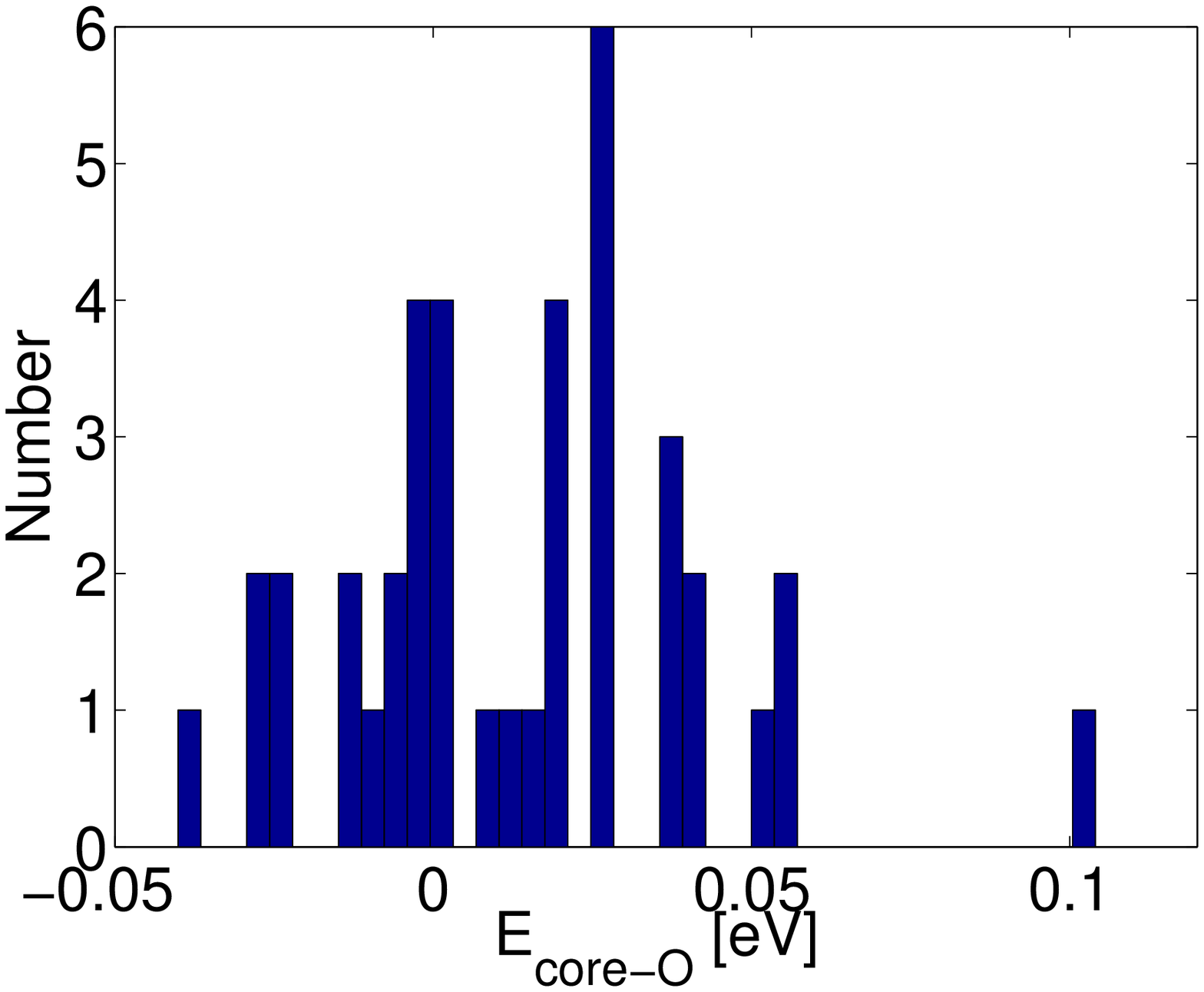}}
 \subfigure[]{\includegraphics[width=0.45\textwidth]{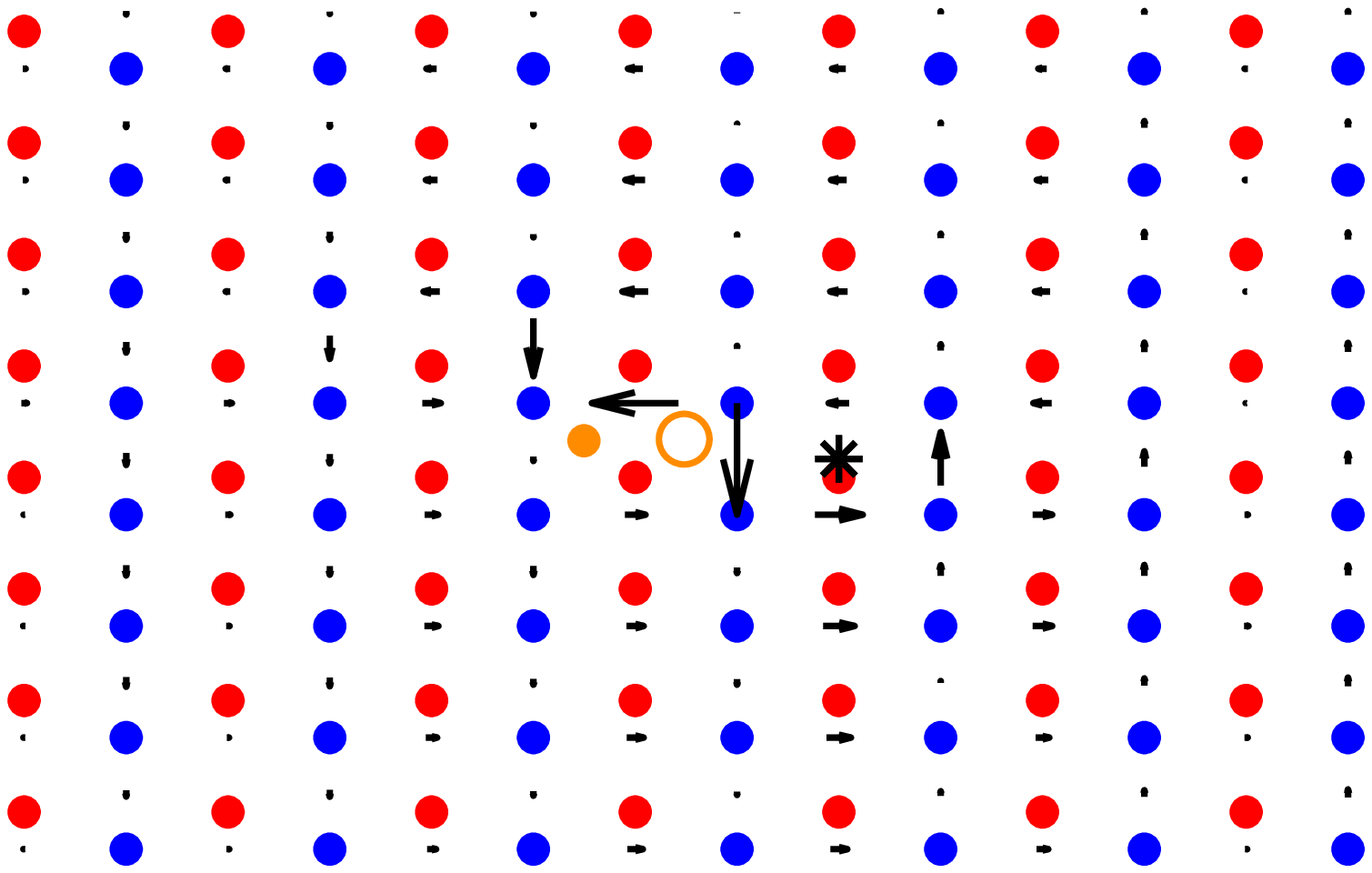}}
 \subfigure[]{\includegraphics[width=0.45\textwidth]{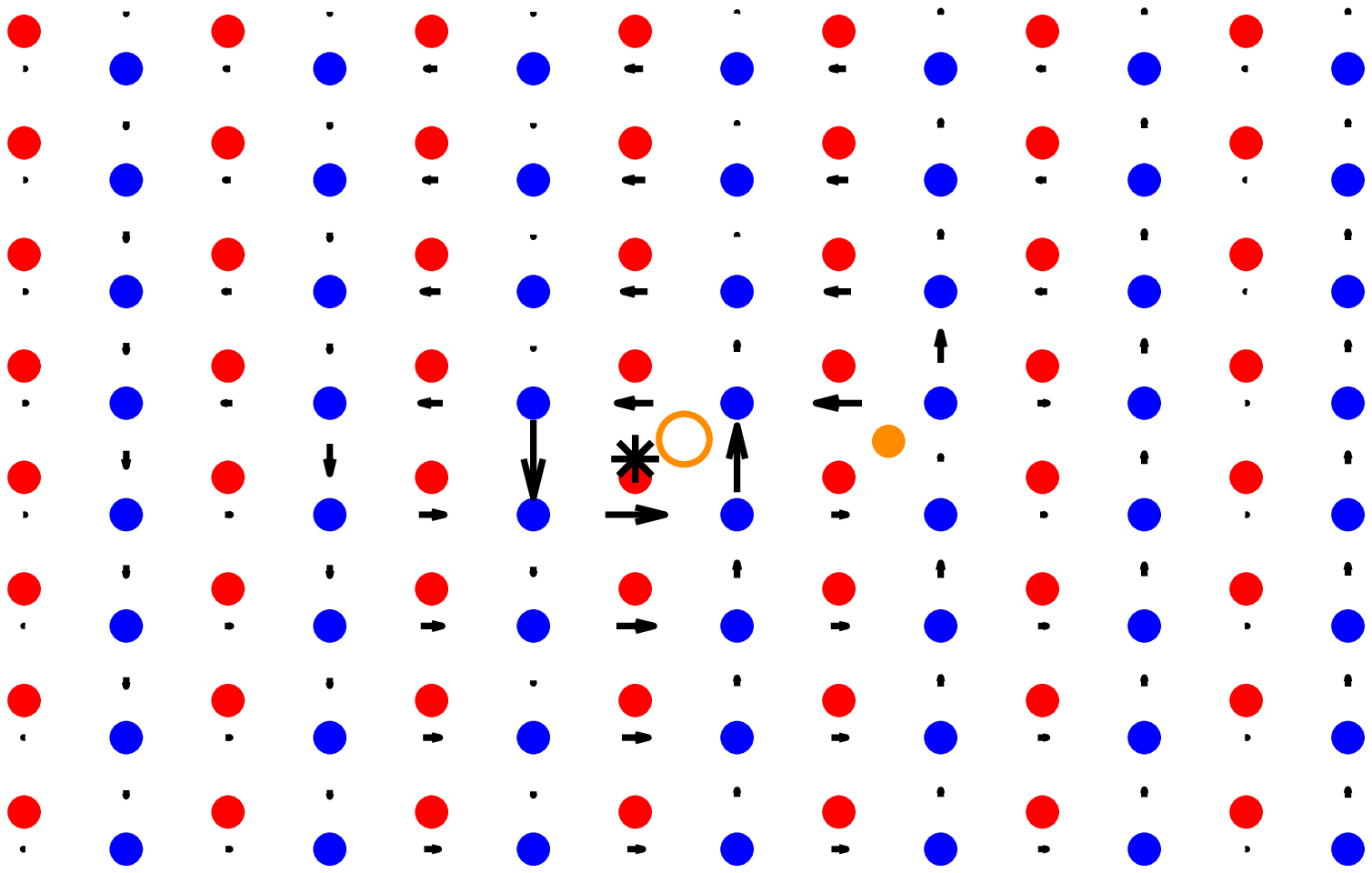}}
 \subfigure[]{\includegraphics[width=0.45\textwidth]{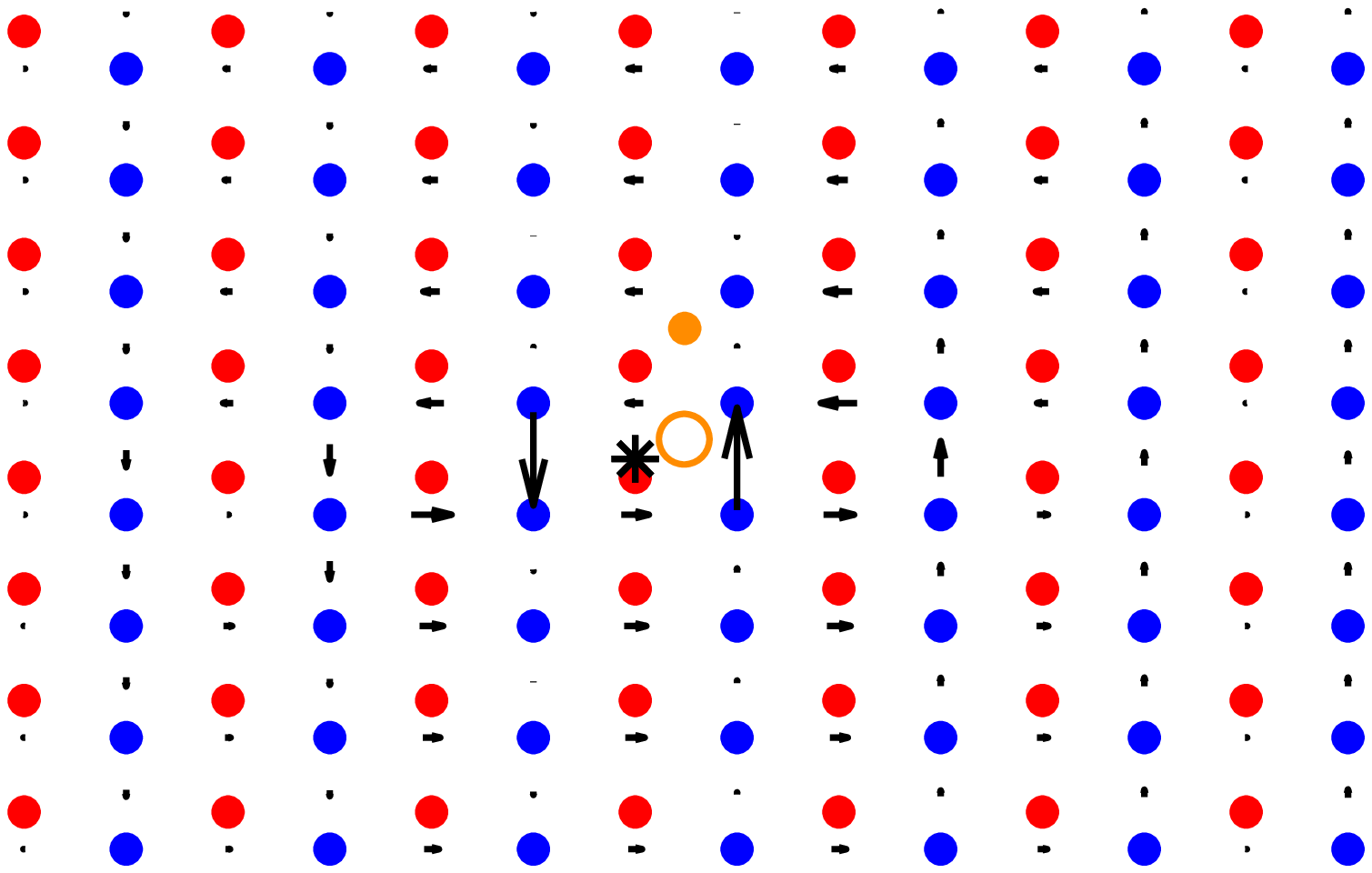}}
 \caption{(a) Histogram of interaction energies between asymmetric/symmetric dislocation core and oxygen atom at various positions. (b)-(d): DD maps between only type A Ti atoms for oxygen at several typical positions. The meaning of each symbol is the same as Fig. \ref{fig_core_ss} (d). The golden circle is located in the same position as the dashed circle in Fig. \ref{fig_core_pure} (f), which can be defined as original dislocation core center. (b): oxygen is $\frac{\langle c \rangle}{2}$ away from the original center. (c): oxygen is $\langle c \rangle$ away from the original center. (d): oxygen is at the nearby prismatic plane with zero distance along $\langle c \rangle$ from the original center.}
 \label{fig_inter_E}
\end{figure}

The histogram of oxygen solute-dislocation interaction energies for all 40 configurations (asymmetric and symmetric) is plotted in Fig. \ref{fig_inter_E} (a). It clearly shows that almost all the interaction energies are relatively small, and on the scale of $\pm$ 0.06 eV, which is comparable to $k_{\textup{B}}$ $\times$ T = 0.0259 eV, at T = 300 K. So there are relatively weak interactions between the oxygen and the $\langle a \rangle$ screw dislocation core, as long as the oxygen is not located inside the region with large lattice displacements. There is one exception, with repulsive energy of $\sim$ 0.1 eV, corresponding to the case that an oxygen interstitial is $\frac{\langle c \rangle}{2}$ away from the center octahedral site; the corresponding DD map is shown in Fig. \ref{fig_inter_E} (b). We can see that the position of the dislocation core center, the primitive cell with one integrated Burgers vector, is shifted away from the original position labeled with the golden open circle for pure Ti in Fig. \ref {fig_core_pure} (f). According to Fig. \ref{fig_core_pure} (e) and (f), the interstitial volume of the octahedral sites $\frac{\langle c \rangle}{2}$ away from the center octahedral site is also largely reduced and their shapes are also largely distorted, indicating that there could be also significant repulsive interaction between the dislocation core and the oxygen interstitial. Thus, the dislocation core center can also be pushed away from the original position as the DD map shown in Fig. \ref{fig_inter_E} (b) indicates.

Fig. \ref{fig_inter_E} (c) and Fig. \ref{fig_inter_E} (d) show DD maps between type A atoms for two other typical cases in all 40 configurations. The former corresponds to the case that the oxygen interstitial is $\langle c \rangle$ away from the original octahedral center at the same prismatic glide plane; the latter is the case where O is at the prismatic plane nearest to the glide plane with 0 distance along $[0001]$ direction to the original octahedral center. Apparently, although the introduction of the oxygen interstitial changes the structures of dislocation cores to a certain extent, the position of the dislocation core center, i.e., the primitive cell with one integrated Burgers vector, does not change. This is consistent with the weak interaction energies shown in Fig. \ref{fig_inter_E} (a) and also the distribution of interstitial volume in Fig. \ref {fig_core_pure} (e) and (f).  The largest changes of interstitial volumes occurs for sites on the same prismatic plane of the dislocation core with the range of $\frac{\langle c \rangle}{2}$ $\sim$ $\langle c \rangle$ away from the original octahedral center. Thus, it can be concluded that the interaction between oxygen and an $\langle a \rangle$-type screw dislocation core is very short ranged ($\sim$ 5 $\textup{\AA}$) and is very directional, being most pronounced for sites on the same prismatic plane as the major glide plane of the $\langle a \rangle$ screw dislocation\cite{Naka88,Yu14}.

\subsubsection{Long-range interactions between oxygen interstitial and $\langle a \rangle$ screw dislocation}
\label{sec_long_range}
The weak interaction between an oxygen interstitial and the screw dislocation for sites outside the small core region is also consistent with the linear elasticity theory. In isotropic elasticity theory, a perfect screw dislocation generates a pure shear stress field, which does not interact with a solute atom that creates simple volumetric strain\cite{Hirth92}. The interaction remains relatively weak (on the scale of 0.1 eV) even when anisotropic elasticity of dislocation and the solute atom are taken into account\cite{Clouet08,Yasi10,Hanlumyuang10,Leyson10}. To check the effect of anisotropy in the present case, we first calculate the stress/strain induced by the occupation of oxygen at an octahedral site in the perfect Ti lattice in a octahedral supercell defined in Sec. \ref{sec_method_strain}. The results show that, besides the volumetric changes, oxygen at octahedral sites only generate shear stress/strain on the $(11\bar{2}0)$ plane perpendicular to $[11\bar{2}0]$. This result can be easily understood by the projections of atomic positions of Ti and O atoms shown in Fig. \ref{fig_cells} (a). It shows that projections of Ti atoms on $(0001)$ plane (the left part of Fig. \ref{fig_cells} (a)) have mirror symmetry relative to $(11\bar{2}0)$ and $(1\bar{1}00)$ across the position of the O atom, and similar mirror symmetry can also be found on $(1\bar{1}00)$ plane. So the chemical bonds between this O atom and all nearby Ti atoms also have such mirror symmetry on these two planes, which produce no shear deformation. But on the $(11\bar{2}0)$ plane (the right part of Fig. \ref{fig_cells} (a)), no mirror symmetry exists relative to $(0001)$ or $(1\bar{1}00)$ across the projection of the O position. On the other hand, in a hcp structure a straight screw dislocation along $[11\bar{2}0]$ generates only shear strain components on $(0001)$ and $(1\bar{1}00)$ without other volumetric or shear components\cite{Chou62,Hirth92}. Thus, by a first-order approximation, there should be no elastic interactions between this screw dislocation and O interstitials located at octahedral sites in a hcp structure.

\subsubsection{Mechanical shuffle of oxygen interstitial}
\label{sec_O_shuffle}

\begin{figure}[th]
\subfigure[]{\includegraphics[width=0.45\textwidth]{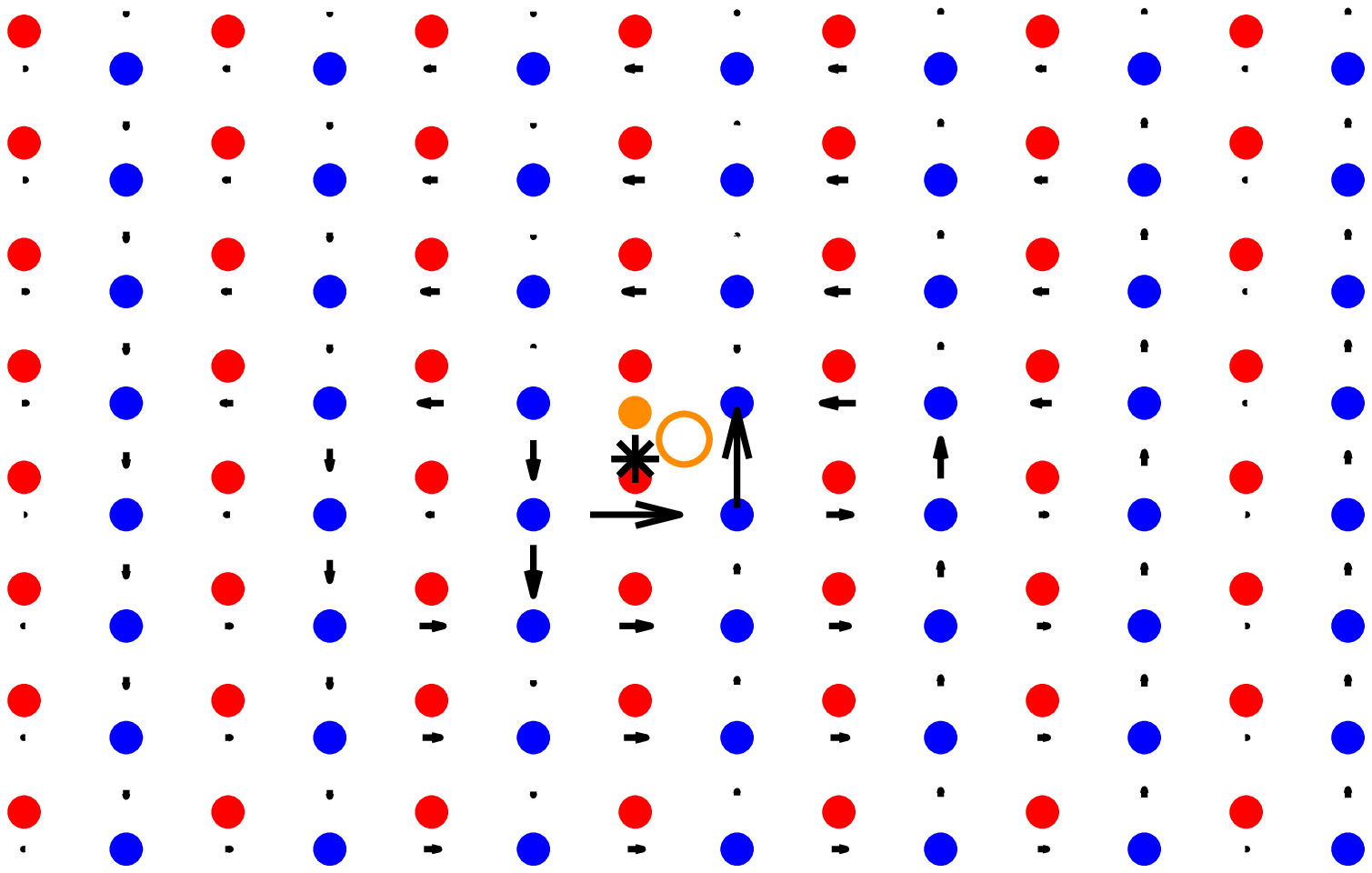}}
 \subfigure[]{\includegraphics[width=0.45\textwidth]{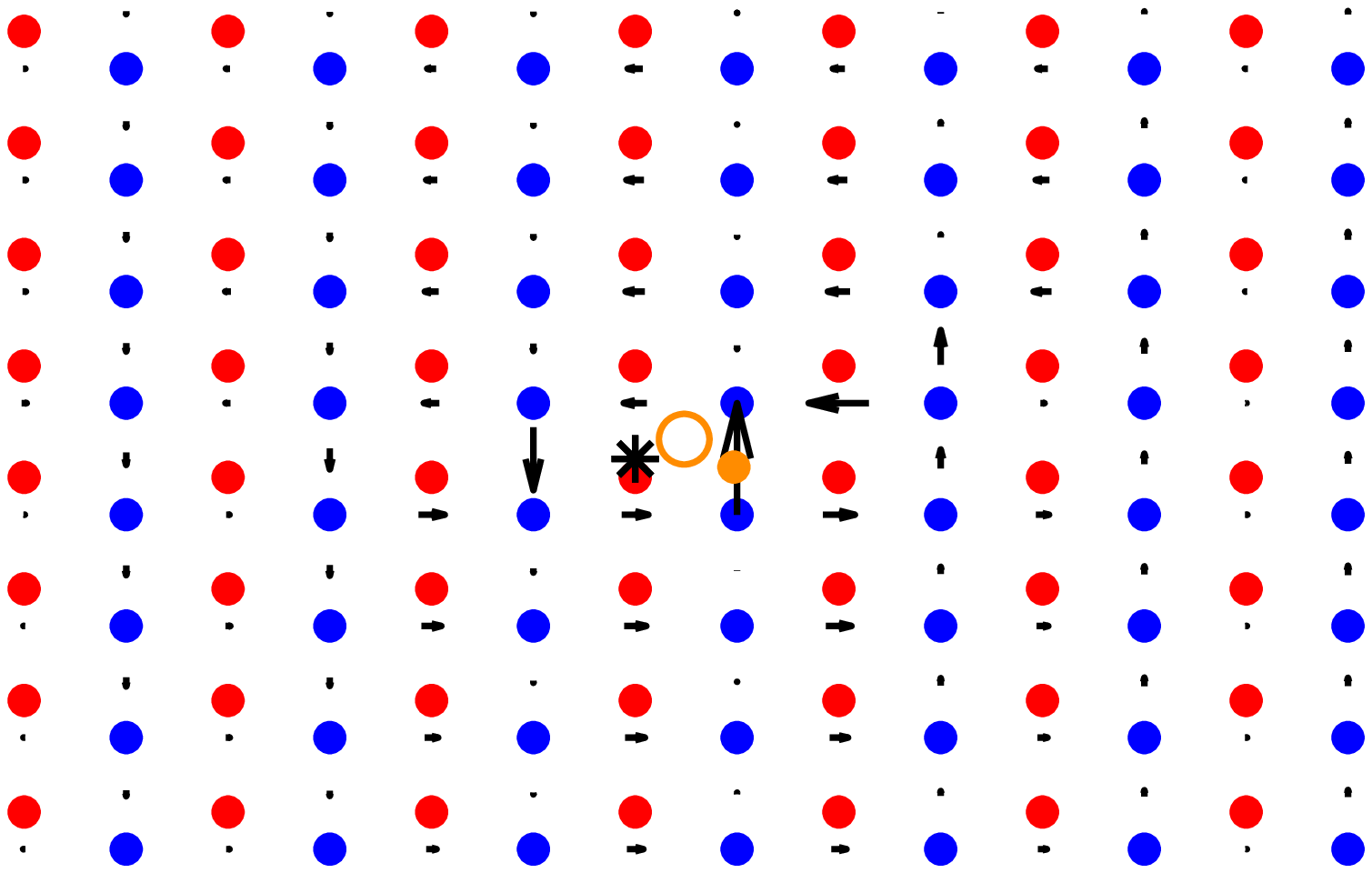}}
 \caption{(a) and (b): DD map between only type A Ti atoms for oxygen at type B (a) and type A (b) basal plane near the original dislocation core center in the symmetric core, respectively. The meaning of each symbol is the same as Fig. \ref{fig_core_ss} (d).}
 \label{fig_core_basal}
\end{figure}

All the above calculations characterize the interaction strength and range of the interactions between an $\langle a \rangle$-type screw dislocation and an oxygen interstitial solute located at its normal octahedral site, which is the most stable site in the undistorted Ti structure. However, as shown in Sec. \ref{sec_GSF}, there could be other stable sites in the dislocation core when there is large misfit across the slip plane, and these sites cannot be reached automatically in the DFT relaxations, due to the large barrier for oxygen to shuffle inside Ti. Thus, instead of the original octahedral sites between two basal planes, the oxygen is also inserted at the new octahedral site on either of two basal planes in the relaxed configuration of the dislocation core, where there is large interstitial volume according to Fig. \ref{fig_core_pure} (e) and (f). Both the asymmetric and symmetric cores are investigated and the relaxed DD maps for the symmetric core case are shown in Fig. \ref{fig_core_basal}. Similarly to those for Al substitution in Fig. \ref{fig_core_ss} (b), although the internal atomic structures change, the position of the dislocation core center determined by the primitive cell with one Burgers vector integrated displacement does not change (this is also found for the asymmetric core, not shown here). In addition, compared with oxygen far away from the dislocation core, it only costs 0.05$\sim$0.10 eV extra energy to insert oxygen into the basal plane of the dislocation core. From the GSF energy calculations in Sec. \ref{sec_GSF}, $E^{\textup{int}}_{\textup{O-Path II}}(x=0.5a)$ = 0.14 eV in (2$\langle a \rangle$ $\times$ 1$\langle c \rangle$) supercell and converge to 0.10 eV in (3$\langle a \rangle$ $\times$ 2$\langle c \rangle$) supercell. These results suggest the repulsive energies estimated from direct dislocation core-interstitial interactions are on the same scale of its approximate value $E^{\textup{int}}_{\textup{O}}(x)$ derived from GSF energy calculations. The differences come from the inhomogeneous distribution of lattice displacements inside the dislocation core and the lateral interactions between oxygen in the image supercells along $\langle a \rangle$ direction, which is also close to 0.1 eV according to previous DFT calculations\cite{Ruban10}. All these structural and energetic results suggest that it is likely that oxygen interstitials can exist in a stable form at such new octahedral sites on the basal plane in the dislocation core, which is not possible in the undistorted structure of $\alpha$-Ti due to the close-packed structure of basal plane. In fact, such oxygen interstitial structures were observed by high-resolution scanning-transmission electron microscopy (HR-STEM) in $\alpha$-Ti\cite{Yu14}, which is a strong evidence to support the validity of our first-principles simulations.

\section{Summary and Discussion}
\label{sec_sum}

In summary, we have employed first-principles DFT calculations to investigate the interaction between Al and O solute atoms and [11$\bar{2}$0] $\langle$ a $\rangle$-type screw dislocations in $\alpha$-Ti. Both GSF energy and dislocation core calculations show that, when compared with a substitutional solute atom like Al, a large interstitial atom like O has a much stronger but short-ranged and directional repulsive interaction with the $\langle a \rangle$-type screw dislocation. This strong repulsive interaction for the interstitial case can even lead to the mechanical shifting of an oxygen solute from an original octahedral site to a new position on the basal plane in the dislocation core. This strong interaction qualitatively explains the large solid solution strengthening effects of oxygen to pure Ti found in experiments.  While the above calculations are performed for oxygen interstitials, many of the results correlate with the changes in interstitial volumes induced by the dislocation, such that we expect qualitatively similar results for large interstitial solutes like C and N, more generally.

In our simulations, because of the high oxygen concentration of interstitial solutes arising from the small supercell size along the $\langle a \rangle$ direction, the dislocation core can be pushed away as a whole straight line when it is close to oxygen at the octahedral site. However, in a realistic case when the oxygen concentration is small, other plastic deformation mechanisms are much more energetically favorable. One possible scenario is that only local parts of the screw dislocation line near the oxygen interstitial cross slips to the nearby prismatic plane. In this case, the segments of dislocation lines that connect the two parts of the screw dislocations on two nearby prismatic plane can be quite immobile\cite{Yu14}, which may even act as two pinning points as a Frank-Read sources\cite{Hirth92}. Another possible mechanism is that the dislocation core pushes the oxygen interstitial away from the original octahedral site to the new site on the basal plane. Then as the dislocation core sweeps away, the oxygen interstitial atom has two possible choices. It can either jump back to their original octahedral site if the perfect lattice is almost recovered, or it can be pushed away to next octahedral site in front of the dislocation line. Under large plastic deformation where there are many screw dislocations gliding on the prismatic plane, this multiple-step oxygen shuffle could be a possible mechanism for oxygen interstitial atoms to accumulate in front of certain screw dislocation lines, which would further impede the mobility of the dislocations. Thus, large-scale simulations, based on the basic interaction mechanisms obtained from this paper, should be applied to quantitatively understand the strengthening effects and toughness degradations for oxygen and other large interstitial atoms in $\alpha$-Ti.

\begin{acknowledgments}

We gratefully acknowledge funding from the U.S. Office of Naval Research under grant N00014-12-1-0413.  T.T. acknowledges financial support of the Japanese Ministry of Education, Culture, Sports, Science and Technology (MEXT), Grant-in-Aid for Scientific Research in Innovative Areas ``Bulk Nanostructured Materials.Ó

\end{acknowledgments}

\bibliography{MyBibliography}

\clearpage

\end{document}